\documentclass[11pt,twoside]{article}

\usepackage{fullpage}
\usepackage{subcaption}

\usepackage{epsf}
\usepackage{fancyheadings}
\usepackage{graphics}
\usepackage{graphicx}
\usepackage{psfrag}
\usepackage{pgfmath}
\usepackage{enumitem}

\usepackage{bbm, dsfont}

\usepackage{color}
\usepackage{amsthm}
\usepackage{amsfonts}
\usepackage{amsmath}
\usepackage{amssymb}
\usepackage{algorithm}
\usepackage{algorithmic}

\usepackage{hyperref}   
\hypersetup{colorlinks,allcolors= blue}

\usepackage{appendix}
\usepackage{titletoc}
\usepackage{minitoc}
\usepackage{adjustbox}
\usepackage{caption}
\usepackage{booktabs, multirow} 
\usepackage{soul}
\usepackage[table]{xcolor} 
\usepackage{changepage,threeparttable} 
\usepackage{natbib}
\usepackage{apalike}

\theoremstyle{plain}

\newtheorem{theos}{Theorem}
\newtheorem{props}{Proposition}

\newtheorem{remark}{Remark}

\newtheorem{definition}{Definition}

\newlength{\widebarargwidth}
\newlength{\widebarargheight}
\newlength{\widebarargdepth}

\makeatletter
\long\def\@makecaption#1#2{
        \vskip 0.8ex
        \setbox\@tempboxa\hbox{\small {\bf #1:} #2}
        \parindent 1.5em  
        \dimen0=\hsize
        \advance\dimen0 by -3em
        \ifdim \wd\@tempboxa >\dimen0
                \hbox to \hsize{
                        \parindent 0em
                        \hfil 
                        \parbox{\dimen0}{\def\baselinestretch{0.96}\small
                                {\bf #1.} #2
                                } 
                        \hfil}
        \else \hbox to \hsize{\hfil \box\@tempboxa \hfil}
        \fi
        }
\makeatother

\long\def\comment#1{}

\newcommand{\Exs}{\ensuremath{{\mathbb{E}}}}
\newcommand{\Prob}{\ensuremath{{\mathbb{P}}}}

\newcommand{\NORMAL}{\ensuremath{\mathcal{N}}}

\newcommand{\Dspace}{\ensuremath{\mathcal{D}}}
\newcommand{\R}{\mathbb{R}}

\newcommand{\X}{\ensuremath{\mathcal{X}}}
\newcommand{\Y}{\ensuremath{\mathcal{Y}}}
\newcommand{\Z}{\ensuremath{\mathcal{Z}}}

\newcommand{\data}{\mathcal{D}}

\newcommand{\Id}{\mathbf{I}_d}

\newcommand{\lb}{\left \{}
\newcommand{\rb}{\right \}}

\newcommand{\lp}{\left(}
\newcommand{\rp}{\right )}

\newcommand{\mrd}{\mathrm{d}}

\newcommand{\zpool}{\Dspace^{\mathrm{P}}}

\newcommand{\ngroup}{\ensuremath{K}}
\newcommand{\nk}{\ensuremath{n_k}}
\newcommand{\nktr}{\ensuremath{\tilde{n}_k}}
\newcommand{\Xki}{\ensuremath{X_{k,i}}}

\newcommand{\tXi}{\ensuremath{\widetilde{X}_i}}
\newcommand{\hXi}{X_i^{\mathrm{H}}}
\newcommand{\hYi}{Y_i^{\mathrm{H}}}

\newcommand{\tX}{\ensuremath{\widetilde{X}}}
\newcommand{\btX}{\ensuremath{\widetilde{\mathbf{X}}}}

\newcommand{\bpX}{\ensuremath{\mathbf{X}^\prime}}

\newcommand{\tYi}{\ensuremath{\widetilde{Y}_i}}

\newcommand{\tY}{\ensuremath{\widetilde{Y}}}

\newcommand{\Yki}{\ensuremath{Y_{k,i}}}
\newcommand{\PXk}{\ensuremath{P^{\scalebox{0.6}{(k)}}_{X}} }
\newcommand{\PX}{\ensuremath{P^{}_{X}} }
\newcommand{\PXone}{\ensuremath{P^{\scalebox{0.6}{(1)}}_{X}} }
\newcommand{\PXtwo}{\ensuremath{P^{\scalebox{0.6}{(2)}}_{X}} }
\newcommand{\PbtX}{P_{\btX}}
\newcommand{\PbpX}{P_{\bpX}}

\newcommand{\PM}{\ensuremath{ \widetilde{P}_{X} }}

\newcommand{\PS}{\ensuremath{ P^{\star}_{\phixz} }}

\newcommand{\PY}{\ensuremath{P^{}_{Y|X}}}
\newcommand{\PYZ}{\ensuremath{P^{}_{Y|Z}}}

\newcommand{\score}{s}

\newcommand{\Xz}{\ensuremath{X_{0}}}

\newcommand{\Yz}{\ensuremath{Y_{0}}}

\newcommand{\dz}{\mathrm{dim}(\mathcal{Z})}
\newcommand{\sth}{\sigma^2_{\mathrm{h}}}

\newcommand{\kK}{k \in [\ngroup]}

\newcommand{\ncal}{n}

\newcommand{\pwi}[1]{p^w_i(#1)}
\newcommand{\pwmi}[1]{p^{\Bar{w}}_i(#1)}

\newcommand{\pwz}[1]{p^w_0(#1)}
\newcommand{\pwmz}[1]{p^{\bar{w}}_0(#1)}
\newcommand{\pwkz}[1]{p^{w_k}_0(#1)}
\newcommand{\pwki}[1]{p^{w_k}_i(#1)}

\newcommand{\wk}{w_k}

\newcommand{\wm}{\Bar{w}}

\newcommand{\wh}{\hat{w}}

\newcommand{\whe}{\hat{w}^{\mathrm{adj}}} 
\newcommand{\ws}{w^\star}

\newcommand{\model}{\mathcal{M}}

\newcommand{\zktr}{\ensuremath{\Dspace^{\scalebox{0.6}{(k)}}_{\mathrm{tr}}} }
\newcommand{\zkcal}{\Dspace^{\scalebox{0.6}{(k)}}_{\mathrm{cal}}}
\newcommand{\zk}{\Dspace^{\scalebox{0.6}{(k)}}}

\newcommand{\zztr}{\Dspace^{\scalebox{0.6}{(0)}}_{\mathrm{tr}}}
\newcommand{\ztrall}{\Dspace^{\scalebox{0.6}{(0:K)}}_{\mathrm{tr}}}

\newcommand{\phixz}{\Phi}
\newcommand{\supp}{\mathrm{supp}}

\newcommand{\QZ}{\ensuremath{Q^{}_{\phixz}} }
\newcommand{\PZk}{\ensuremath{P^{\scalebox{0.6}{(k)}}_{\phixz} }}
\newcommand{\PZM}{\ensuremath{ \widetilde{P}_{\phixz} }}

\newcommand{\QX}{\ensuremath{Q^{}_{X}} }

\newcommand{\zcalh}{\mathcal{D}^{\mathrm{H}}}
\newcommand{\zcalhone}{\mathcal{D}_{(1)}^{\mathrm{H}}}
\newcommand{\zcalhtwo}{\mathcal{D}_{(2)}^{\mathrm{H}}}

\newcommand{\Conf}[1]{\mathcal{C}(#1)}

\newcommand{\Confk}[1]{\mathcal{C}^{\scalebox{0.6}{(k)}}(#1)}

\newcommand{\Confm}[1]{\mathcal{C}^{\mathrm{M}}(#1)}
\newcommand{\Confwm}[1]{\mathcal{C}^{\mathrm{WM}}(#1)}
\newcommand{\Confdp}[1]{\mathcal{C}^{\mathrm{P}}(#1)}

\newcommand{\quan}[1]{\mathrm{Quantile}\left( #1\right)}

\newcommand{\wpvaluek}[1]{p_k(#1)}
\newcommand{\wpvalueone}[1]{p_1(#1)}
\newcommand{\wpvalueK}[1]{p_{\ngroup}(#1)}
\newcommand{\wpvaluem}[1]{p_{\mathrm{merge}}(#1) }

\newcommand{\indic}[1]{\mathbbm{1}_{ #1 }}
\newcommand{\inde}[1]{\mathbbm{1}_{ \{#1\} }}

\newcommand{\htauk}{\hat{\tau}_k}
\newcommand{\Confhh}[1]{\mathcal{C}^{\mathrm{H}}(#1)}

\newcommand{\pS}{p_{\star}}
\newcommand{\hpS}{\hat{p}_{\star}}
\newcommand{\hpSe}{\hat{p}_{\star}^{\mathrm{adj}}} 
\newcommand{\htau}{\hat{\tau}}
\newcommand{\tauk}{\tau_k}
\newcommand{\taumin}{\tau_{\min}}
\newcommand{\taumax}{\tau_{\max}}
\newcommand{\Cw}{C_w}
\newcommand{\calA}{\mathcal{A}}

\newcommand{\epsn}{\epsilon_n}
\newcommand{\err}{\mathrm{Err}}
\newcommand{\erra}{\err_{\mathcal{A}}}
\newcommand{\errac}{\err_{\mathcal{A}^c}}

\begin{document}

\doparttoc 
\faketableofcontents 
\part{} 


\begin{center}
{\bf{\Large{Uncertainty Quantification With Multiple Sources}}}

\vspace*{.2in}

{\large{
\begin{tabular}{ccc}
Mufang Ying$^{\dagger}$, Wenge Guo$^{\ddagger}$, Koulik Khamaru$^{\dagger}$, Ying Hung$^{\dagger}$ \\
\end{tabular}
}}

\vspace*{.2in}

\begin{tabular}{c}
Department of Statistics, Rutgers University - New Brunswick$^\dagger$\\
Department of Mathematical Sciences, New Jersey Institute of Technology$^\ddagger$\\
\end{tabular}

\vspace*{.2in}

\today

\vspace*{.2in}

\begin{abstract}
Weighted conformal prediction (WCP) has been commonly used to quantify prediction uncertainty under covariate shift. However, the effectiveness of WCP relies heavily on the degree of overlap between the training and test covariate distributions. This challenge is exacerbated in multi-source settings with varying covariate distributions, where direct application of WCP can be impractical. In this paper, we address the multi-source setup by leveraging WCP under the assumption of a shared conditional distribution. We investigate two extensions of WCP: (i) a merge-based aggregation of source-specific weighted conformal prediction sets, and (ii) a data-pooling strategy that jointly reweights samples across all sources. Theoretical guarantees are provided for the proposed approaches, and experiments are conducted based on a synthetic regression task and a multi-domain image classification benchmark to validate our proposed methods.
\end{abstract}

\end{center}

\section{Introduction}
Consider a diagnostic setting where pathologists at a target medical center aim to provide reliable uncertainty quantification for tumor diagnosis, specifically in the form of prediction sets.  Because the target center lacks a sufficient repository of annotated biopsy samples, it seeks to leverage labeled archives from multiple other hospitals. This approach operates under the assumption that the fundamental pathological characteristics of malignancies remain consistent across institutions. A critical hurdle, however, is that the data distributions often differ substantially due to heterogeneous imaging protocols (e.g., scanner resolution, staining techniques) and varying patient demographics. Incorporating data from these diverse sources introduces significant challenges related to covariate shift, a pervasive issue in medical domain adaptation~\citep{sun2015survey,quinonero2022dataset}. This naturally raises the question: how can such multi-source data be effectively utilized to construct valid uncertainty estimates for the target patient population?

Conformal prediction, a methodology for constructing prediction intervals, has gained significant attention and popularity for the ability to assess uncertainties with machine learning models~\citep{vovk1999machine,papadopoulos2002inductive,vovk2005algorithmic,lei2013distribution,lei2015distribution,angelopoulos2023conformal}. One of the reasons for the prominence of conformal prediction is its capacity to provide nonasymptotic coverage guarantees for any black box algorithms that remain unaffected by the underlying distribution. This remarkable feature is achieved by relying on the exchangeability of the data points. A recent framework, weighted conformal prediction~\citep{tibshirani2019conformal}, addresses the covariate shift setting by incorporating knowledge of the likelihood ratio between the training and test covariate distributions. 

\begin{figure}[!t]
    \centering
    \begin{subfigure}[b]{0.3\textwidth}
        \centering
        \includegraphics[width=\textwidth]{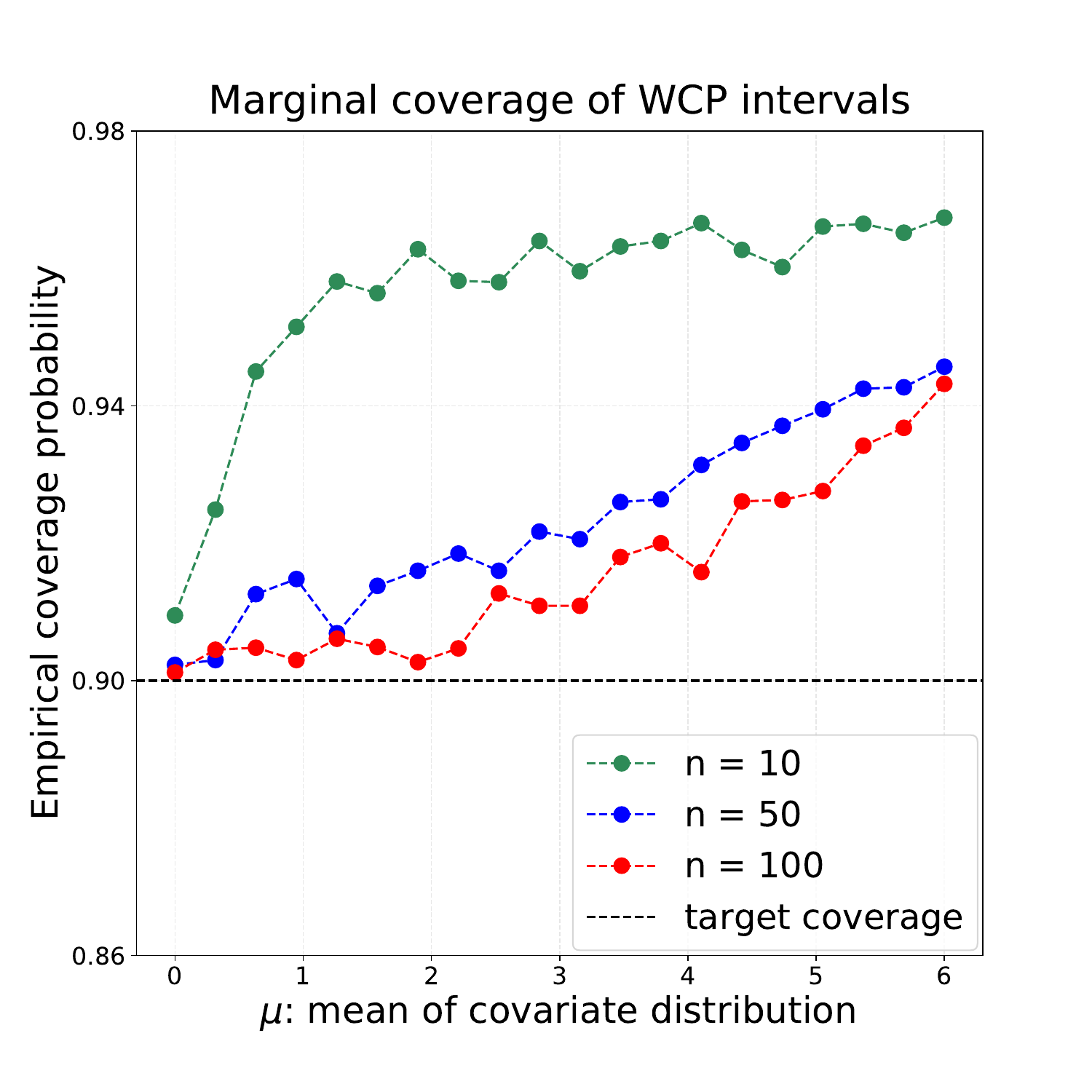}
    \end{subfigure}
    \begin{subfigure}[b]{0.3\textwidth}
        \centering
        \includegraphics[width=\textwidth]{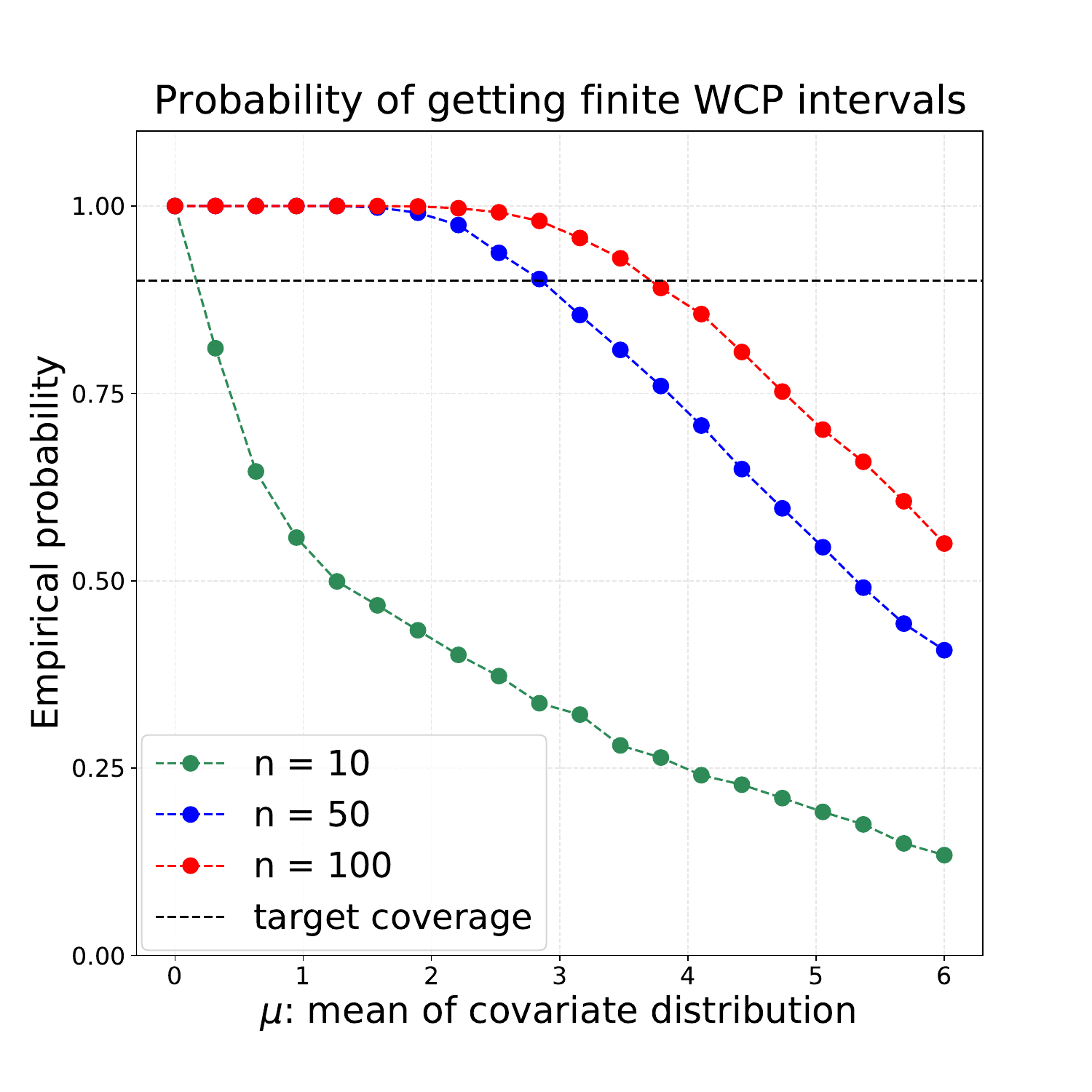}
    \end{subfigure}
    \begin{subfigure}[b]{0.3\textwidth}
        \centering
        \includegraphics[width=\textwidth]{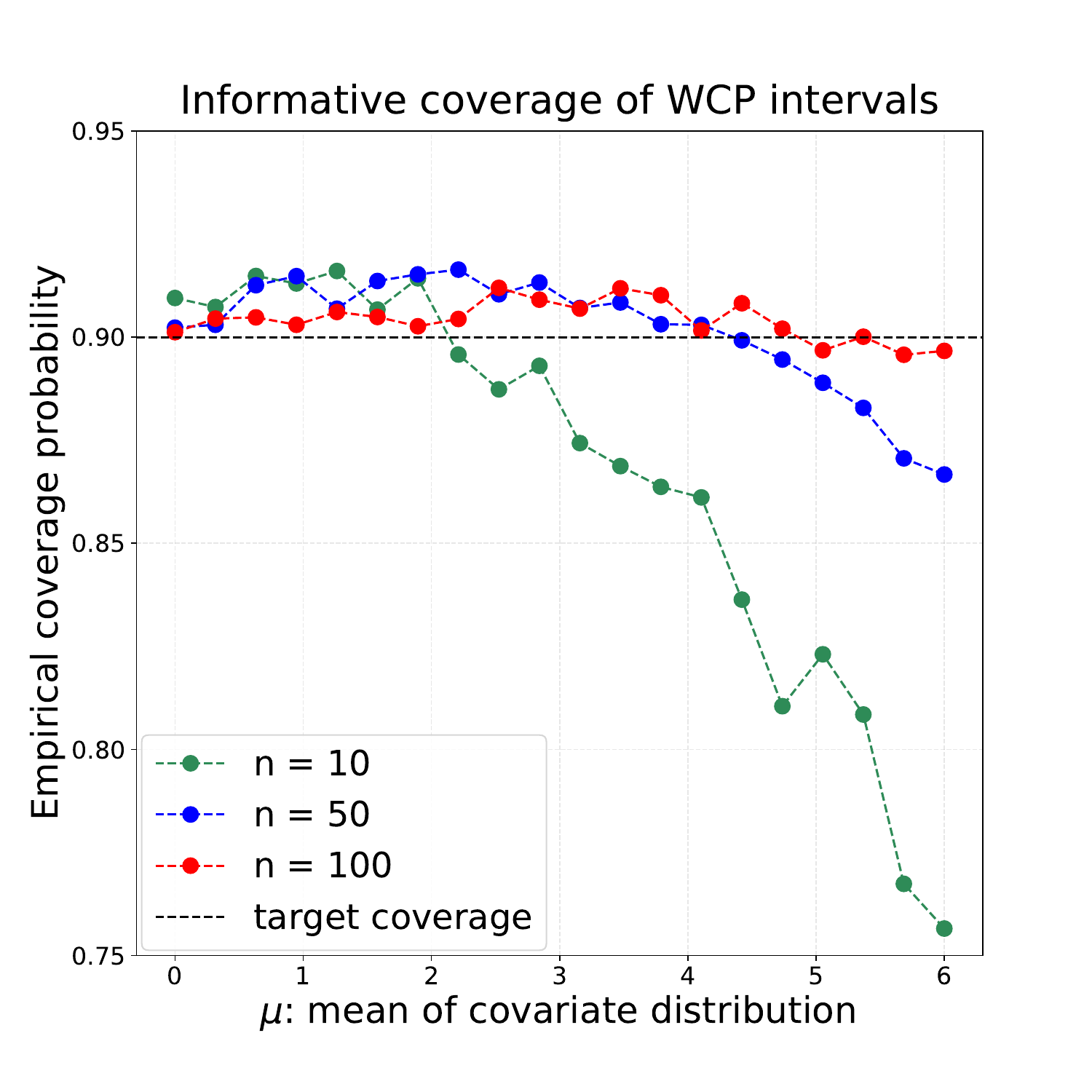}
    \end{subfigure}
    \caption{ An application of weighted conformal prediction to $\PY=\NORMAL( \text{sigmoid}(X), 0.01)$, $\QX = \NORMAL(0,9)$ and $\PX = \NORMAL(\mu,9)$ with mean $\mu \in [0, 6]$ at level $0.9$. \textbf{Left:} empirical marginal coverage probability. \textbf{Middle:} empirical probability of getting finite prediction intervals. \textbf{Right:} empirical conditional coverage given prediction interval is finite. Sample size $n \in \{10,50,100\}$ are considered and the results are obtained through $10000$ replications. See Appendix for additional details and another example with varying variance of $\PX$.}
    \label{fig:fix_var}
\end{figure}

While weighted conformal prediction has demonstrated successful applications in diverse domains such as experimental design, survival analysis and causal inference (e.g., see~\citet{fannjiang2022conformal,lei2021conformal,candes2023conformalized}), the effectiveness of this framework relies on the overlap of covariate distributions between training and test.  In Figure~\ref{fig:fix_var}, a simple example is used to demonstrate that the constructed WCP intervals can be $(-\infty, \infty)$ in certain cases. We examine a regression example with $\QX = \NORMAL(0,9)$ representing the covariate distribution of test data and $\PX = \NORMAL(\mu,9)$ representing the covariate distribution of training data. Three sample sizes, $n=10, 50, 100$ are considered and the empirical results for the constructed WCP intervals are obtained from $10000$ replications with absolute residual being non-conformity score function. The overlap of covariate distributions $\PX$ and $\QX$  is reduced when increasing $\mu$ - mean of $\PX$. Although WCP intervals provide marginal coverage above the target level $0.9$ (the left panel of Figure~\ref{fig:fix_var}), this coverage guarantee is accomplished at the cost of an increasing probability of uninformative prediction intervals, $(-\infty, \infty)$, as the overlap decreases (the middle panel of Figure~\ref{fig:fix_var}). Furthermore, conditioning on finite prediction intervals, it is shown on the right panel of Figure~\ref{fig:fix_var} that the conditional coverage decreases and falls below target level as the overlap decreases. The decrease of the conditional coverage appears to be more significant for smaller sample size. In high-dimensional settings such as image classification, however, the assumption of covariate overlap in the original input space is almost never satisfied due to the curse of dimensionality. Consequently, the support of the source distribution often has negligible or zero intersection with the support of the target distribution, which leads to density ratio estimation unstable or undefined. As a result, applying weighted conformal prediction directly on raw images often yields degenerate importance weights, leading to prediction sets that are either empty or trivially contain all possible classes. Furthermore, extending the generalized WCP techniques of \citet{tibshirani2019conformal} to the multi-source setting raises significant practical challenges. In particular, it remains unclear how to effectively integrate data from multiple heterogeneous domains within the standard WCP framework. The corresponding weight functions, as defined under the generalized weighted conformal prediction approach in \cite{tibshirani2019conformal}, can become prohibitively complex in the multi-source scenario, limiting their practical applicability.

In this paper, we investigate the multi-source setting under the assumption that the conditional distribution is shared across domains. To address covariate shift in high-dimensional regimes, we introduce the concept of a domain-invariant sufficient transformation, which maps high-dimensional inputs (e.g., raw images) into a lower-dimensional representation that preserves the relevant covariate information while facilitating overlap across domains. Under this transformation, the covariate shift structure is inherited from the original feature space. Building on this framework, we propose two principled extensions of weighted conformal prediction: (i) a merge-based aggregation approach that combines source-specific weighted conformal prediction sets, and (ii) a data-pooling strategy that jointly reweights samples from all sources to construct a weighted prediction set. We establish theoretical guarantees for both methods.

\subsection*{Related work}
Existing literature includes work that assumes exchangeability or symmetry between the target and source covariate distributions~\citep{dunn2023distribution,lee2023distribution,duchi2024predictive,dobriban2025symmpi}. For instance, the hierarchical model in~\citet{dunn2023distribution} assumes that the covariate distributions of the source domains and the target domain are exchangeable, treating them as i.i.d. draws from a common super-population. In contrast, our work does not make this assumption, as the target domain may exhibit a unique covariate distribution that cannot be treated as similar to the source domains. A recent work~\citet{bhattacharyya2024group} focuses on achieving a marginal coverage guarantee in a special scenario where both training and test data can be viewed as collected via stratified sampling. Specifically, the covariate $X$ is represented as $X = (X^0, X^1)$, with $X^0 \in [\ngroup]$ encoding the domain information, and the test distribution exhibits covariate shift only at $X^0$. Their approach is compatible with our setting when the target covariate distribution can be expressed as a mixture of source distributions with known weights; however, as our objective is to develop WCP-based methods with provable coverage guarantees, we avoid limiting our investigation to this specific scenario. Finally, \citet{liu2024multi} address multi-source conformal prediction under general distribution shifts; however, their methodology requires partial labeled data from the target domain and operates outside the WCP framework, distinguishing it from our setting and objective.

\section{Problem formulation}
\label{sec:multi-group}

As discussed in the previous section, our primary focus is predictive uncertainty quantification involving multiple source domains and a single target domain. Let $\zk = \{(\Xki,\Yki) \in \X \times \Y: i\in [\nk + \nktr] \}$ denote the dataset collected from the $k$-th domain for $k \in [\ngroup]$. Observations from the $k$-th domain are assumed to be independent and identically distributed (i.i.d.) according to:
\begin{equation}\label{eq:group-distri}
(\Xki, \Yki) \stackrel{\mathrm{i.i.d}}{\sim} \PXk\times \PY \qquad \text{for $i \in [\nk]$},
\end{equation}
where $\PXk$ denotes the marginal covariate distribution in the $k$-th domain. We assume a shared conditional distribution $\PY$ across all groups, and that observations are independent across groups. Adopting the split conformal prediction framework~\citep{vovk2005algorithmic,papadopoulos2002inductive}, we partition the dataset $\zk$ into two disjoint sets: a training set $\zktr = \{(\Xki,\Yki) : i \in [\nk + \nktr] \setminus [\nk] \}$, and a calibration set $\zkcal = \{(\Xki,\Yki) : i \in [\nk] \}$.

For the target group, we denote the joint distribution as $\QX \times \PY$, where $\QX$ represents the marginal covariate distribution in the target domain, and the target conditional distribution $\PY$ remains identical to that of the source domains. Note that the covariate distribution $\QX$ typically differs from the covariate distributions of the source domains. We assume that an unlabeled dataset $\zztr$ for the target domain is available. Let $(X_0, Y_0)$ denote an independent test data point sampled from the target distribution $\QX \times \PY$, with $Y_0$ unobserved. Our objective is to construct a prediction set $\Conf{X_0}$ for this unknown outcome $Y_0$ that satisfies the following marginal coverage guarantee:
\begin{equation*}
    \Prob \left\{ \Yz \in \Conf{\Xz} \right\} \geq 1 - \alpha,
\end{equation*}
where $\alpha \in (0,1)$ is the pre-specified miscoverage level.

\subsection*{Role of $\PY$ and domain-invariant sufficient transformation}
Let us consider a scenario where the supports of the covariate distributions across different sources are mutually disjoint. In such regimes, labeled data from source domains fails to provide empirical evidence for the conditional distribution of $Y$ given $X$ within the target domain, leading to the fact that uncertainty quantification in the target domain is fundamentally unidentifiable without additional structural assumptions.

To make the problem well-posed, we restrict our analysis to a subclass of problems where the dependence of $Y$ on $X$ is mediated by a latent representation. 
\begin{definition}
    Let $\phixz:\X \to \Z$ denote a measurable function from the raw covaraite space $\X$ to a latent feature space $\Z$. We define $\phixz$ as a domain-invariant sufficient (DIS) transformation for the response variable $Y$ if it satisfies the following two conditions:
    \begin{enumerate}
        \item[] \textbf{Sufficiency} Given the latent features $Z = \phixz(X)$, response variable $Y$ is conditionally independent of the raw covariates $X$ given the latent features $Z = \phixz(X)$:
        \begin{equation*}
        Y \;\perp\; X \mid Z, \quad \text{i.e.,} \quad \Prob(Y \mid X) = \Prob(Y \mid Z).
    \end{equation*}

        \item[]  \textbf{Representation support overlap}   The support of the latent representation in the target domain is contained in the union of the supports of the latent representations in the source domains, i.e.,
        \begin{equation} \label{eq:latent-overlap}\supp(\QZ) \subseteq \bigcup_{k=1}^K \supp(\PZk),
        \end{equation}
        where $\QZ$ and $\PZk$ denote the marginal distributions of $Z = \phixz(X)$ in the target and $k$-th source domain, respectively.
        
    \end{enumerate} 
\end{definition}
\noindent The first condition also known as conditional independence condition, implies that $\phixz(X)$ is a sufficient statistic for $Y$ - the raw covariates $X$ contain no additional predictive information about $Y$ that is not already captured by $Z$. The second condition ensures that the support of target domain is covered by the aggregate source domains in the latent feature space, which enables effective knowledge transfer even when the individual source domains have disjoint supports. We point out that by employing a DIS transformation, the covariate shift problem can be effectively mapped from the original space $\mathcal{X}$ to the latent feature space $\mathcal{Z}$, where the distributional overlap is guaranteed. Throughout the paper, we assume the DIS transformation $\phixz$ is available.

\paragraph{Example on classification}
Suppose that $Y = \psi(X)$ for some measurable function 
$\psi : \mathcal{X} \to \mathcal{Y}$ (i.e., there is no label noise and $\psi$ is the ground truth labeling function). In this case, sufficiency condition in DIS transformation is equivalent to the existence of a 
measurable function $g : \mathcal{Z} \to \mathcal{Y}$ such that
\[
Y = g(\phixz(X)) \quad \text{almost surely}.
\]
Thus, $\phixz$ meets the sufficiency condition if and only if the label function $\psi$ can factor through $\phixz$. Therefore, if $\phixz$ is too coarse (e.g., a constant map), then $Y$ cannot be recovered from $\phixz(X)$. If $\phixz$ preserves the label-relevant information, then it is sufficient.

Lastly, we introduce some additional notations. Let $\ztrall= \cup_{k\in \{0\} \cup [\ngroup]} \zktr $ be the aggregated training dataset, which includes data from both source and target domains and is used for likelihood ratio estimation or model training.
For regression tasks with overlapping covariate distributions, $\phixz$ can be the identity map, whereas for image classification, $\phixz$ may be a feature extractor, either learned from data or based on pre-trained models. We let $\model$ denote the predictor built on top of $\phixz$. We denote by  $s: \X \times \Y \to \R$ a non-conformity score function measuring the discrepancy between the observed response and the model prediction. Common examples include the absolute residual, $s(x,y) = |y - \model(\phixz(x))|$, for regression, and $s(x,y) = 1 - \model(\phixz(x))_y$ for classification, where $\model(\phixz(x))$ is the predicted probability vector.

\newcommand{\PZ}{P^{}_{\phixz}}

\section{Weighted conformal prediction with multiple sources}
First, we review the weighted conformal prediction framework~\citep{tibshirani2019conformal} in the context where a sufficient representation is available. Building on this foundation, we then discuss methods for leveraging weighted conformal prediction to address uncertainty quantification for the target domain with multiple sources. 

\subsection{Weigted conformal prediction}
Following Theorem 2 in~\cite{tibshirani2019conformal}, we have the following result:
\begin{props}
\label{prop:wcp}

Let $\data = \{ (X_i,Y_i) \in \X\times \Y: i\in [n]\}$ denote a dataset of $n$ i.i.d. observations drawn from the source distribution $\PX \times \PY$. Let $(\Xz,\Yz)$ be an independent test point drawn from the target distribution $\QX \times \PY$. Suppose that $\phixz$ is a DIS transformation for the response variable $Y$. Furthermore, assume the likelihood ratio $w(z) = \mathrm{d} \QZ(z) / \mathrm{d} \PZ(z)$ is known, where $\QZ$ and $\PZ$ denote the push-forward distributions of $\QX$ and $\PX$ under $\Phi$, respectively. We define the weight function at any $x \in \X$ as follows:

\begin{equation}
\label{eq:wcp-pwi}
   \pwz{x;\data } = \frac{ w(\phixz(x))}{ w(x) + \sum_{j=1}^n w(\phixz(X_j))}   \quad\text{and} \quad \pwi{x;\data } = \frac{ w(\phixz(X_i)) }{ w(\phixz(x) + \sum_{j=1}^n w(\phixz(X_j))}, 
\end{equation}
where $i\in[n]$. Then, for $\alpha \in (0,1)$, it follows that
\begin{equation}
\label{eq:wcp-result}
  \Prob \lb  \Yz \in  \Conf{\Xz;1-\alpha} \rb  \geq 1-\alpha,
\end{equation}
where
\begin{align*}
\label{eq:wcp-conf}
\scalebox{1}{
 $\Conf{x;1-\alpha} = \lb y:  s(x,y) \leq \quan{1-\alpha; \sum_{i = 1}^n \pwi{x;\data}  \delta_{\score(X_i, Y_i)} + \pwz{x;\data} \delta_{\infty}} \rb$} \\
 \text{and  $\delta_{z}$ denotes a unit point mass at $z\in\R$}.
\end{align*}
\end{props}

By reweighting the non-conformity scores on dataset $\mathcal{D}$, one can achieve finite-sample coverage guarantees. However, such validity does not imply efficiency. Note that, while equation~\eqref{eq:wcp-result} ensures the required marginal coverage for $Y_0$, the resulting prediction sets can be excessively conservative, as demonstrated in Figure~\ref{fig:fix_var}. This conservatism is particularly pronounced in cases of severe covariate shift - for instance, when the total variation distance between the source and target distributions, defined as $\mathrm{d}_{\mathrm{TV}}(\PZ,\QZ) = \sup_{A}|\PZ(A) - \QZ(A)|$, is large.

While weighted conformal prediction establishes a theoretical foundation for handling covariate shift, extending it to the multi-source setting presents significant challenges, such as determining how to combine heterogeneous datasets. In the following sections, we introduce methods specifically designed to leverage multiple source domains effectively.Throughout our theoretical analysis, we assume that a DIS transformation $\phixz$ for the response variable is available. By applying the weighted conformal prediction procedure outlined in Proposition~\ref{prop:wcp} to these extracted latent features, we can build valid prediction sets for any individual source.

\subsection{Merged weighted conformal prediction}
A natural baseline approach to the multi-source problem is to aggregate the weighted conformal prediction sets from all available domains. To formalize this, we first consider the the prediction set obtained from the individual source domain $k$. With known likelihood ratio $\wk = \mrd \QZ / \mrd\PZk$ under $\phixz$,  the prediction set can be constructed as
\begin{align*}
\scalebox{0.9}{
$\Confk{x; 1-\alpha} = \lb y: \score(x,y) \leq  \quan{1-\alpha; \sum_{i = 1}^{\nk} \pwki{x;\zkcal}  \delta_{\score(\Xki, \Yki)} + \pwkz{x;\zkcal} \delta_{\infty}}  \rb$}.
\end{align*}
According to Proposition~\ref{prop:wcp}, for each $k\in [K]$, we have 
\begin{equation}
\label{eq:K-wcp}
    \Prob \lb \Yz \in \Confk{\Xz; 1-\alpha} \rb \geq 1-\alpha.
\end{equation}
Given these valid single-source sets, merging them via a voting mechanism emerges as the most natural strategy. Following Theorem 2.4 in~\citet{gasparin2024merging}, we can construct a merged $(1-\alpha)$-level prediction set at $x$ as follows:
\begin{equation}
\label{eq:merge-wcp}
\Confm{x; 1 - \alpha, \gamma} = \lb y: \frac{1}{K}\sum_{k = 1}^{K} \inde{ y \in \Confk{x;1-(1 - \gamma)\alpha}} > \gamma \rb,
\end{equation}
for any parameter $\gamma \in [0,1)$. This prediction set includes all values $y$ that are included in at least $\lceil \gamma \ngroup\rceil$ of the adjusted WCP sets from $\ngroup$ source domains. This formulation encompasses several classical combinations: when $\gamma = 1/2$, the procedure corresponds to the majority vote procedure~\citep{cherubin2019majority,solari2022multi,gasparin2024merging}; conversely, when $\gamma = (K-1)/K$, the procedure becomes equivalent to Bonferroni's procedure. We summarize this result in the following proposition for Merged WCP:
\begin{props}
Given merged prediction bands constructed in equation~\eqref{eq:merge-wcp}, it follows that 
\begin{equation}
    \Prob \lp \Yz \in \Confm{\Xz; 1 - \alpha, \gamma}   \rp \geq 1 - \alpha.
\end{equation}
\end{props}
In the appendix, we present a more generalized result connecting merging weighted prediction sets via merging weighted conformal p-values~\citep{barber2025unifying,jin2023model}. We comment that while the merging strategy provides robust coverage, it tends to be statistically conservative, often yielding prediction sets that are larger than necessary due to the loose bounds inherent in voting or Bonferroni-type procedure.

\subsection{Pooled weighted conformal prediction}
In contrast to the merged weighted conformal prediction approach, which aggregates prediction sets from distinct domains, data pooling through weighted conformal prediction combines all sources into a single composite dataset. By effectively leveraging the total sample size across all domains, pooling offers the potential for tighter, more statistically efficient intervals compared to the often conservative voting methods. However, while pooling is a standard practical approach, its validity under heterogeneous covariate shifts is not automatically guaranteed. This section aims to investigate the theoretical validity of the pooling strategy.

The data pooling procedure can be performed by permuting the elements in $\cup_{k\in[K]} \zkcal$ uniformly at random to obtain $\zpool$
    \begin{equation*}
        \zpool =  \lb (\tX_1,\tY_1) ,\ldots, (\tX_n, \tY_n) \rb \quad \text{with } n = \sum_{k\in [K]} \nk.
    \end{equation*}
One can easily verify that  for any $i\in[n]$
\begin{equation}
\label{eq:weighted-pop}
      (\tXi, \tYi)\sim \PM \times \PY \qquad \text{where} \quad \PM =  \sum_{\kK} \frac{\nk}{n} \PXk.
\end{equation}
Note that, after the data permutation step, observations share a covariate distribution $\PM$ as indicated in equation~\eqref{eq:weighted-pop}, enabling the utilization of the WCP framework. However, the marginal coverage guarantee of WCP breaks apart due to the correlation within the dataset $\zpool$. Nonetheless, in cases where correlations within $\zpool$ under $\phixz$ are minimal, the difference between $\zpool$ under $\phixz$ and its i.i.d. version is insignificant.  To summarize, we have the following theorem for Pooled WCP:
\begin{theos}
\label{theo:pool}
Let $\PZM$ denote the push forward distribution of $\PM$ under DIS transformation $\phixz$. Then, it follows that
\begin{equation}
\label{eq:wcp-pool}
    \Prob \lb \Yz \in  \Confdp{\Xz;1-\alpha, \zpool} \rb  \geq 1- \alpha - \mathrm{d}_{\mathrm{TV}} ( \phixz(\btX),  \phixz(\bpX) ),
\end{equation}
where $\wm = \mrd \QZ/ \mrd \PZM$,
\begin{align*}
     \scalebox{0.88}{  $\Confdp{x;1-\alpha, \zpool}  = \lb y\in \R:  s(x,y) \leq \quan{1-\alpha; \sum_{i = 1}^n \pwmi{x;\zpool}  \delta_{\score(\tXi, \tYi)} + \pwmz{x;\zpool} \delta_{\infty}} \rb$,} \\
      \text{and} \; \scalebox{0.88}{ $X_i^\prime \stackrel{\mathrm{i.i.d}}{\sim} \PM$, $\phixz(\btX) = (\phixz(\tX_1^{\prime})^\top, \ldots, \phixz(\tX_n^{\prime})^\top )$,  $ \phixz(\bpX) = (\phixz(X_1^{\prime})^\top, \ldots, \phixz(X_n^{\prime})^\top )$} .
\end{align*}
Moreover, if the latent feature distributions $\PZk$ are domain-invariant (i.e., identical across all source domains), we have
\begin{equation*}
      \Prob \lb \Yz \in  \Confdp{\Xz;1-\alpha, \zpool} \rb  \geq 1- \alpha.
\end{equation*}
\end{theos}
\noindent Equation~\eqref{eq:wcp-pool} ensures that Pooled WCP yields almost valid coverage when the joint covariate distribution $\phixz(\btX)$ is close to its i.i.d. counterpart $\phixz(\bpX)$ in TV distance. For a deeper theoretical analysis of exchangeable mixtures and their i.i.d. counterparts, see, e.g., \citet{ehm1991binomial,han2024approximate}. We comment that assuming $\PZk$ is invariant across domains is a strong condition, as it  can implicitly constrain both the structure of $\phixz$ and the label distributions. Nevertheless, the resulting coverage lower bound~\eqref{eq:wcp-pool} remains quite conservative; see appendix for an example.

\begin{remark}
In practice, we emphasize that, when estimating the likelihood ratio using pooled data from source domains, the proportion of samples drawn from each source domain should match the relative sizes of their corresponding calibration datasets.
\end{remark}

\paragraph{Hierarchical structure for source domains}
To address the uncontrolled coverage bound in equation~\eqref{eq:wcp-pool}, we alternatively provide an analysis for data pooling through WCP under a two-level hierarchical structure, which achieves asymptotic coverage guarantee. Specifically, the process for generating the $i$-th sample is modeled as follows:
\begin{enumerate}
\item First, a source domain index $k_i$ is selected from $\{1, \dots, K\}$ according to a \text{Multinomial} distribution with unknown positive probabilities $(\tau_1, \dots, \tau_K)$.
\item Second, given a selected domain $k_i$, a data pair $(\hXi, \hYi) $ is drawn from the joint distribution $\PXk \times \PZ$.
\end{enumerate}
Here, we use the notation $\zcalh = \{ (k_i, \phixz(\hXi), \hYi ) \}_{i\in [\ncal]}$ for the dataset under the hierarchical model and the DIS transformation $\phixz$. By marginalizing out the domain index, the resulting data points can be viewed as i.i.d. draws from the mixture distribution:
\begin{equation}\label{eq:H-pop}(\phixz(\hXi), \hYi) \stackrel{\mathrm{i.i.d}}{\sim} \PS \times \PYZ \qquad \text{where} \quad \PS = \sum_{\kK} \tau_k \PZk.
\end{equation}
Therefore, we randomly partition $\zcalh$ into two disjoint subsets, $\zcalhone$ and $\zcalhtwo$. For the first subset, $\zcalhone$, we discard the group indices $\{k_i\}$ and work only with the data pairs $(\phixz(\hXi), \hYi)$, which will serve as our calibration set. Conversely, from the second subset, $\zcalhtwo$, we use only the indices $\{k_i\}$ to estimate the unknown mixture weights, yielding $\{\htauk\}$. By applying weighted conformal prediction using estimated mixture weights, we obtain the asymptotic coverage gaurantee:

\begin{theos}
    \label{theo:poolh}
Let $\wh = \mrd \QZ/ \mrd (\sum \htauk \PZk)$ and assume the likelihood ratio $\ws = \mrd \QZ/\mrd \PS$ is bounded, i.e., $\sup_z \ws(z) \le \Cw < \infty$.
For $k\in [\ngroup]$, let  $\beta_k = \frac{\max(\htau_k, \epsn)}{ \sum_{j \in [K]} \max(\htau_j, \epsn) }  $ where $\{\htau_k\}_{k\in[\ngroup]}$ are the estimated mixture weights using $\zcalhtwo$, and $\epsn = 1/|\zcalhtwo|$. Define the adjusted likelihood ratio estimator $\whe = \mathrm{d} \QZ / \mathrm{d} (\sum_{k} \beta_k \PZk)$,  then it follows 
    \begin{equation}
    \label{eq:poolh}
         \Prob \lb \Yz \in  \Confhh{\Xz;1-\alpha, \zcalhone} \rb  \geq 1 - \alpha  - O\lp \sqrt{\frac{K}{|\zcalhtwo|}}\rp,
    \end{equation}
where
\begin{align*}
\scalebox{0.8}{
 $\Confhh{x;1-\alpha, \zcalhone}  = \lb y\in \R:  s(x,y) \leq \quan{1-\alpha; \sum_{(\hXi,\hYi) \in \zcalhone} p_i^{\whe}(x;\zcalhone)  \delta_{\score(\hXi, \hYi)} + p_0^{\whe}(x;\zcalhone) \delta_{\infty}} \rb$}.
\end{align*}
\end{theos}
Note that the coverage gap results from the estimation error associated with the parameters $\{\tauk\}_{k\in[\ngroup]}$. To keep this error bounded, we employ the adjusted likelihood ratio described in the theorem above. We also point out that the analysis of this estimation error mirrors the analysis in Section 3.2 of \citet{bhattacharyya2024group}. However, our setting is more general: unlike their work, we do not restrict the target covariate distribution to be a simple mixture of the source covariate distributions.

\section{Experiments}
To compare the empirical performance of the aforementioned methods, we consider two experimental settings: a synthetic regression task and an image classification task using Digit-Five dataset. In both scenarios, the primary goal is to construct valid prediction sets for a target source by utilizing calibration data gathered from $K$ distinct heterogeneous sources. Across the experiments, we consider the following methods: Conformal Prediction, Pooled WCP, and Merged WCP with $\gamma = 1/2$ and $\gamma = (K-1)/K$.

\subsection{Synthetic regression task}
In this experiment, we simulate a synthetic regression task where the true underlying response surface is non-linear and depends on the first two coordinates of the feature vector. Specifically, the response is generated as $Y = f(X) + \sigma(X)\cdot \epsilon$ with $\epsilon \sim \NORMAL(0,1)$, where the mean function $f(X)$ and the noise scale $\sigma(X)$ are defined as:
\begin{equation*}
    f(X) = \frac{1}{1 + \exp( |X_1|/2 )} \cdot \frac{1}{1 + \exp( |X_2|/2)}, \quad \text{and}\quad \sigma(X) = 1.0 + \frac{\sth}{1 + \|X\|_2}.
\end{equation*}
Here, $\|\cdot\|_2$ is the Euclidean norm of the covariates, and $\sth \in \{0,4\}$ is a parameter controlling the noise variance. Setting $\sth = 4$ introduces strong input-dependent noise, makeing the regression problem highly heteroscedastic.

The source domain covariate distributions are simulated from multivariate Gaussians, $\PXk = \NORMAL(\boldsymbol{\mu_k}, \sigma_k^2\Id)$, where $\Id$ is the $d$-dimensional identity matrix. To induce covariate shift, we randomly sample the distribution parameters for each source domain per replication: the means $\boldsymbol{\mu_k} \sim \text{Uniform}[-1, 1]^d$ and the standard deviations $\sigma_k \sim \text{Uniform}[0.5, 1]$. However, the target covariate distribution is fixed to a zero-mean Gaussian $\NORMAL(\mathbf{0}, 0.5\Id)$. We vary the feature dimensionality $d \in \{2, 5, 10\}$ and the number of sources $K \in \{5,10\}$ in the experiments to test the robustness of the methods.

For the predictive modeling and conformal setup, we train a Gaussian Process (GP) regressor with a radial basis function kernel on the combined labeled training splits from all source datasets. We adopt absolute residuals as the non-conformity scores and, following \cite{tibshirani2019conformal}, estimate the likelihood ratios using Random Forest classifiers. Table~\ref{tab:regr} reports the results averaged over 1000 independent replications at a target level of $\alpha = 0.1$. Performance is evaluated based on the Marginal Coverage Probability (MCP), the Proportion of Finite Intervals (PFI), and the Median Length (MedL) of the finite intervals.

\begin{table}[!htp]
\caption{Regression task: method comparison}
\label{tab:regr}
\centering
\resizebox{0.95\textwidth}{!}{
\begin{tabular}{clccccccccc}
\toprule
& 
& \multicolumn{3}{c}{$d = 2$} 
& \multicolumn{3}{c}{$d = 5$}  
& \multicolumn{3}{c}{$d = 10$} \\
\cmidrule(lr){3-5}
\cmidrule(lr){6-8}
\cmidrule(lr){9-11}
Setup & Method
& MCP & PFI & MedL 
& MCP & PFI & MedL 
& MCP & PFI & MedL \\
\midrule

\multirow{4}{*}{$\sth=0,K = 5$}
&CP &89.5\% &100.0\% &3.31 &89.7\% &100.0\% &3.30 &90.3\% &100.0\% &3.29 \\
&Pooled WCP &89.7\% &100.0\% &3.25 &90.3\% &100.0\% &3.30 &89.7\% &100.0\% &3.31 \\
&Merged WCP($\gamma = 1/2$) &95.5\% &95.2\% &3.92 &95.7\% &84.2\% &4.22 &97.7\% &80.7\% &4.33 \\
&Merged WCP($\gamma = (K-1)/K$) &91.6\% &99.8\% &3.64 &95.2\% &98.0\% &4.14 &97.4\% &94.5\% &4.38 \\
\midrule

\multirow{4}{*}{$\sth=0,K = 10$}
&CP &88.7\% &100.0\% &3.30 &91.0\% &100.0\% &3.29 &88.4\% &100.0\% &3.29 \\
&Pooled WCP &89.2\% &100.0\% &3.26 &90.9\% &100.0\% &3.29 &89.2\% &100.0\% &3.30 \\
&Merged WCP($\gamma = 1/2$) &93.1\% &98.9\% &3.80 &95.6\% &90.9\% &4.11 &96.1\% &85.3\% &4.30 \\
&Merged WCP($\gamma = (K-1)/K$) &92.0\% &99.4\% &3.67 &95.9\% &95.3\% &4.31 &97.9\% &91.3\% &4.78 \\
\midrule

\multirow{4}{*}{$\sth=4,K = 5$}
&CP &86.5\% &100.0\% &9.66 &84.4\% &100.0\% &7.65 &85.3\% &100.0\% &6.55 \\
&Pooled WCP &90.6\% &100.0\% &10.84 &90.7\% &100.0\% &8.93 &90.1\% &100.0\% &7.59 \\
&Merged WCP($\gamma = 1/2$) &94.9\% &95.2\% &13.16 &95.8\% &84.2\% &11.50 &97.7\% &80.7\% &9.74 \\
&Merged WCP($\gamma = (K-1)/K$) &91.2\% &99.8\% &11.81 &94.8\% &98.0\% &10.77 &97.2\% &94.5\% &9.59 \\
\midrule
\multirow{4}{*}{$\sth=4,K = 10$}
&CP &84.2\% &100.0\% &9.64 &85.4\% &100.0\% &7.64 &83.2\% &100.0\% &6.56 \\
&Pooled WCP &89.9\% &100.0\% &10.89 &90.7\% &100.0\% &8.89 &89.0\% &100.0\% &7.59 \\
&Merged WCP($\gamma = 1/2$) &93.0\% &98.9\% &12.61 &96.3\% &90.9\% &11.03 &96.5\% &85.3\% &9.53 \\
&Merged WCP($\gamma = (K-1)/K$) &91.1\% &99.4\% &11.77 &95.5\% &95.3\% &10.97 &96.8\% &91.3\% &10.26 \\ 
\bottomrule
\end{tabular}}
\end{table}

While the conformal prediction baseline achieves valid coverage under constant noise ($\sth = 0$), it consistently under-covers the target distribution in heteroscedastic settings ($\tau = 4$). This breakdown highlights the necessity of addressing covariate shifts for heterogeneous sources. Among the methods proposed to address covariate shifts, Pooled WCP proves to be the most efficient. It successfully achieves near-exact empirical coverage while generating the narrowest prediction intervals. Conversely, while the Merged WCP strictly guarantee valid coverage, they are inherently more conservative and produce wider prediction intervals.

\subsection{Image classification}
\begin{figure}[!htp]
    \centering
    \includegraphics[width=0.65\linewidth]{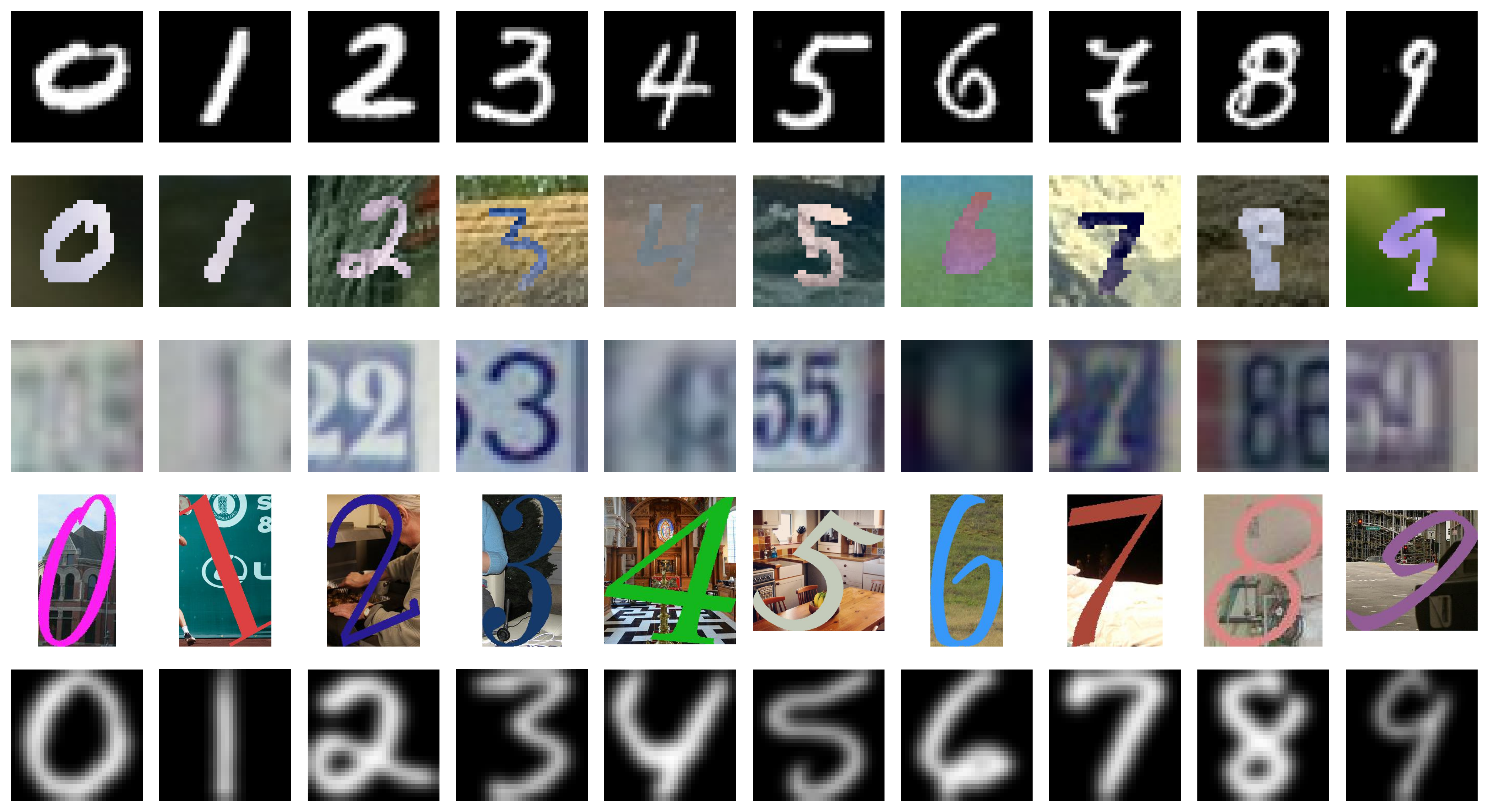}
    \caption{Digit-Five Dataset}
    \label{fig:digits_five}
\end{figure}

For the image classification task, we utilize the standard Digit-Five benchmark~\citep{lecun2002gradient,ganin2015unsupervised}, which comprises five distinct digit recognition datasets: MNIST, MNIST-M, SVHN, USPS, and Synthetic Digits (SYN).  Each dataset represents a different domain with significant visual variations in background, stroke style, and color, thus naturally introducing covariate shift. Recall that the theoretical analysis we provided in previous sections rely on the existence of a DIS transformation. Therefore, to ensure this strong assumption is met in the image classification setting, we utilize an oracle representation - $\phixz$ is learned using data from all sources (including target domain train split) to construct an ideal feature extractor that can serve as DIS transformation. This setup would allow us to isolate the uncertainty quantification problem, and enable us to test the validity and efficiency of the proposed methods.

We consider two training paradigms for a predictive model consisting of a feature extractor $\phixz$ and a classifier $\model$. The feature extractor is implemented as a custom convolutional neural network (CNN). We evaluate the following configurations:
\begin{itemize}
\item \textbf{Base:} Trained end-to-end on the pooled data using the standard cross-entropy loss.
\item \textbf{Base + CL:} Pretrained via contrastive learning, followed by fine-tuning of the classifier.
\end{itemize}
We defer the detailed model architecture and training configuration to the appendix. In the experiment, we vary the dimension of the latent feature space $\mathrm{dim}(\mathcal{Z}) \in \{128,256,512\}$. Since classification accuracies on MNIST and USPS exceed 95\%, whereas those on the remaining datasets are around 90\%, we evaluate performance on the image classification task across three target domains (MNIST-M, SVHN, and SYN) at a target marginal coverage level of $0.95$. The results are summarized in Table~\ref{tab:base_CL}.

\begin{figure}[t]
    \centering
    \includegraphics[width=0.95\linewidth]{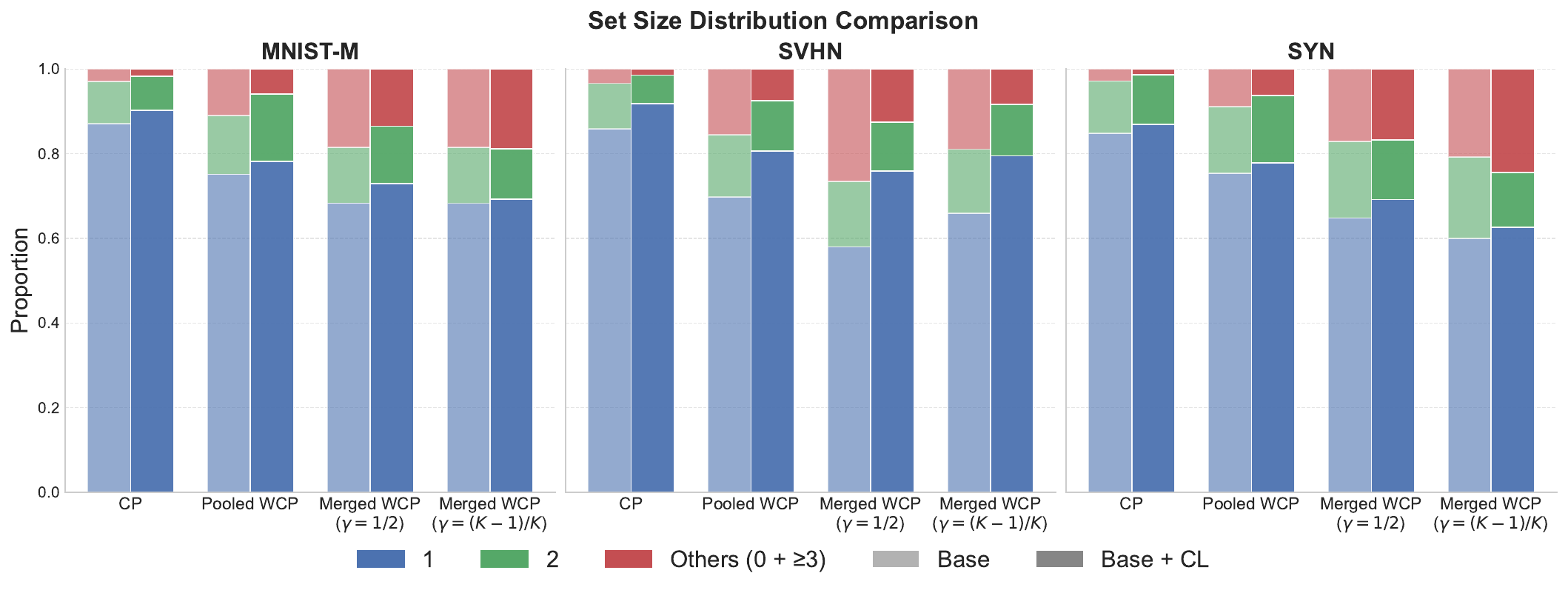}
    \caption{Set size distribution with $\mathrm{dim}(\mathcal{Z}) = 256$ and contrasive learning temperature $\mathrm{temp} = 0.1$.}
    \label{fig:set_dist}
\end{figure}
From Table~\ref{tab:base_CL}, we observe that Vanilla CP is prone to under-coverage under covariate shift. Across multiple target domains, the standard CP baseline frequently fails to attain the nominal coverage level of $0.95$. This under-coverage becomes more pronounced as the latent dimension increases. These results confirm that ignoring covariate shift can lead to unreliable uncertainty quantification.

\begin{table}[b]
\caption{Image classification task: method comparison}
\label{tab:base_CL}
\centering
\resizebox{0.95\textwidth}{!}{
\begin{tabular}{clcccccccccccc}
\toprule
& 
& \multicolumn{4}{c}{MNIST-M}
& \multicolumn{4}{c}{SVHN}
& \multicolumn{4}{c}{SYN} \\
\cmidrule(lr){3-6}
\cmidrule(lr){7-10}
\cmidrule(lr){11-14}
& 
& \multicolumn{2}{c}{Base}
& \multicolumn{2}{c}{Base + CL}
& \multicolumn{2}{c}{Base}
& \multicolumn{2}{c}{Base + CL}
& \multicolumn{2}{c}{Base}
& \multicolumn{2}{c}{Base + CL} \\
\cmidrule(lr){3-4}
\cmidrule(lr){5-6}
\cmidrule(lr){7-8}
\cmidrule(lr){9-10}
\cmidrule(lr){11-12}
\cmidrule(lr){13-14}
$\mathrm{dim}(\mathcal{Z})$ & Method
& MCP & Size
& MCP &  Size
& MCP &  Size
& MCP &  Size
& MCP &  Size
& MCP &  Size \\

\midrule
\multirow{4}{*}{128}
 &CP &93.1\% &1.29 &91.8\% &1.09 &92.3\% &1.27 &91.4\% &1.08 &92.6\% &1.31 &94.2\% &1.17 \\
 &Pooled WCP &96.3\% &1.71 &95.0\% &1.27 &96.5\% &1.79 &94.7\% &1.27 &95.8\% &1.62 &96.5\% &1.41 \\
 &Merged WCP($\gamma = 1/2$) &97.6\% &2.06 &96.7\% &1.52 &98.1\% &2.38 &96.4\% &1.53 &97.3\% &1.94 &98.5\% &1.79 \\
 &Merged WCP($\gamma = (K-1)/K$) &97.2\% &1.95 &93.9\% &1.20 &98.5\% &2.67 &96.7\% &1.59 &98.1\% &2.18 &98.0\% &1.61 \\
\midrule
\multirow{4}{*}{256}
 &CP &89.9\% &1.16 &88.7\% &1.07 &93.1\% &1.18 &91.3\% &1.06 &93.3\% &1.18 &94.2\% &1.13 \\
  &Pooled WCP &94.1\% &1.45 &93.5\% &1.29 &97.0\% &1.63 &95.2\% &1.31 &96.2\% &1.39 &96.3\% &1.30 \\
 &Merged WCP($\gamma = 1/2$) &96.3\% &1.74 &95.5\% &1.55 &98.2\% &2.20 &96.6\% &1.58 &97.8\% &1.70 &98.7\% &1.74 \\
 &Merged WCP($\gamma = (K-1)/K$) &96.4\% &1.74 &97.2\% &1.93 &97.4\% &1.80 &95.5\% &1.35 &98.4\% &1.87 &99.2\% &2.37 \\
\midrule
\multirow{4}{*}{512}
 &CP &85.8\% &1.09 &88.2\% &1.00 &94.6\% &1.22 &92.6\% &1.07 &93.3\% &1.22 &94.8\% &1.11 \\
 &Pooled WCP &93.4\% &1.59 &91.6\% &1.14 &97.7\% &1.70 &95.7\% &1.29 &96.8\% &1.59 &97.7\% &1.39 \\
 &Merged WCP($\gamma = 1/2$) &95.3\% &1.89 &95.8\% &1.42 &97.5\% &1.63 &96.8\% &1.48 &96.7\% &1.61 &98.7\% &1.84 \\
 &Merged WCP($\gamma = (K-1)/K$) &93.0\% &1.56 &92.2\% &1.16 &95.8\% &1.32 &95.9\% &1.34 &98.0\% &1.85 &98.7\% &1.84 \\
\bottomrule
\end{tabular}}
\end{table}
We note that Pooled WCP achieves near-exact coverage and is more efficient when the latent dimension is moderate.
When the latent feature dimension is relatively small ($d \in {128, 256}$), Pooled WCP consistently achieves empirical coverage very close to the target level. For example, under the Base + CL configuration with $\dz=128$, coverage reaches 95.0\% on MNIST-M and 94.7\% on SVHN. Importantly, this is achieved while maintaining relatively small prediction set sizes, demonstrating strong efficiency. On the other hand, Merged WCP is quite conservative. Both merged variants (Merged WCP with $\gamma = 1/2$ and $\gamma = (K-1)/K$) tend to over-cover across domains. While they offer robust validity guarantees, this comes at the cost of noticeably larger prediction sets, suggesting that these methods trade efficiency for stability.

Besides, we see that contrastive pretraining (Base + CL) leads to smaller prediction sets. Across nearly all configurations, models trained with contrastive learning (Base + CL) produce consistently smaller prediction sets compared to the standard Base models, while maintaining comparable coverage. This suggests that contrastive pretraining yields more informative and robust latent representations. A plausible explanation is that improved representation alignment mitigates distributional discrepancies (including mild label shift effects), resulting in sharper predictive uncertainty estimates. Figure~\ref{fig:set_dist} further illustrates this reduction in set sizes for $d=256$. Additional results under different contrastive temperatures are provided in the appendix.

Lastly, we point out that, as the latent dimension increases ($\dz=256$ and especially $\dz=512$), performance degrades for WCP based methods. In particular, for MNIST-M, Pooled WCP fails to meet the target coverage at $\dz=512$ (dropping to 91.6\%). This deterioration is likely due to the curse of dimensionality, which makes likelihood ratio estimation increasingly unstable in high-dimensional spaces. The merged methods are also affected, exhibiting occasional under-coverage.

\section{Conclusion and future work}

In this work, we establish theoretical guarantees for multi-source WCP, there are several limitations and avenues for future exploration. First, standard WCP is inherently bottlenecked in high-dimensional spaces, where likelihood ratio estimation becomes unstable, and weights can degenerate if the target distribution contains modes unseen in the sources. To address this, our analysis relies on the existence of a Domain-Invariant Sufficient (DIS) transformation $\phixz$ to map high-dimensional covariates into a lower-dimensional latent space where likelihood estimation becomes feasible. However, in practice, it is difficult to verify whether a given feature extractor perfectly satisfies the DIS property. Our current theoretical framework relies on this idealized assumption, and consequently, our coverage guarantees do not account for $\epsilon$-approximation errors. It would be interesting to investigate relaxed coverage bounds for scenarios where $\phixz$ is only approximately invariant or partially loses predictive sufficiency. Empirically, given the emergence and representational power of foundation models, one promising future direction is to utilize these large pre-trained models as feature extractors to approximate the DIS property and evaluate our proposed multi-source WCP methods in more complex tasks. Lastly, although we have established marginal coverage validity~\citep{guan2023localized,hore2025conformal,gibbs2025conformal}, it would be valuable to study stronger notions of validity, such as conditional coverage. Another promising direction is to extend the conformal risk control framework to multi-source settings~\citep{angelopoulos2022conformal}, enabling more flexible risk-aware guarantees beyond marginal coverage.

\newpage
\bibliographystyle{apalike}
\small
\bibliography{ref}


\newpage
\appendix

\addcontentsline{toc}{section}{Appendix} 
\part{Appendix} 
\parttoc{}

\section{Sufficiency condition in the example of classification}

Consider the classification setup where $\Y$ is a discrete space (e.g., $\{1, \dots, C\}$).
Suppose $Y = \psi(X)$ almost surely for some measurable function $\psi : \X \to \Y$.
Let $\phi : \X \to \mathcal{Z}$ be measurable and define $Z = \phi(X)$. We aim to show the following are equivalent:

\begin{enumerate}
    \item[(i)] $Y \perp X \mid Z$.
    \item[(ii)] There exists a measurable function $g : \mathcal{Z} \to \Y$ such that
    \[
    \psi(X) = g(\phi(X)) \quad \text{almost surely}.
    \]
\end{enumerate}

\subsection*{(ii) $\Rightarrow$ (i)}
Suppose $Y = g(\phi(X))$ almost surely. Since $Z = \phi(X)$, we have $Y = g(Z)$ almost surely.
Because $Y$ is a deterministic function of $Z$, its conditional distribution given $(X, Z)$ degenerates to a point mass at $g(Z)$. For any class $y \in \Y$:
\[
\Prob (Y = y \mid X, Z) = \Exs[\mathbb{I}(Y=y) \mid X, Z] = \mathbb{I}(g(Z) = y).
\]
Notice that the right-hand side depends only on $Z$. Therefore, marginalizing out $X$ yields the same result:
\[
\Prob (Y = y \mid Z) = \Exs[ \mathbb{I}(g(Z) = y) \mid Z ] = \mathbb{I}(g(Z) = y).
\]
Comparing the two equations, we have:
\[
\Prob (Y = y \mid X, Z) = \Prob (Y = y \mid Z) \quad \forall y \in \Y.
\]
This equality of conditional probabilities for all $y$ implies conditional independence, i.e., $Y \perp X \mid Z$.

\subsection*{(i) $\Rightarrow$ (ii)}
Assume $Y \perp X \mid Z$. By definition of conditional independence, for any class $y \in \Y$,
\[
\Prob (Y = y \mid X, Z) = \Prob (Y = y \mid Z) \quad \text{almost surely}.
\]
Using the premise that $Y = \psi(X)$ is deterministic given $X$, the term on the left is simply an indicator function:
\[
\Prob (Y = y \mid X, Z) = \Exs[\mathbb{I}(Y=y) \mid X, Z] = \mathbb{I}(\psi(X) = y).
\]
Substituting this back into the independence condition:
\[
\mathbb{I}(\psi(X) = y) = \Prob  (Y = y \mid Z).
\]
Let $q_y(z) := \Prob (Y = y \mid Z = z)$. The equation above states that for every $y$, the probability $q_y(\phi(X))$ must be either 0 or 1 almost surely since it equals an indicator function. Also, note that $\sum_{y \in \Y} q_y(z) = 1$, for almost every $z$, there is exactly one class $y^*$ such that $q_{y^*}(z) = 1$, while $q_y(z) = 0$ for all $y \neq y^*$. We can therefore define the function $g: \mathcal{Z} \to \Y$ as:
\[
g(z) = \underset{y \in \Y}{\mathrm{argmax}} \ \Prob(Y = y \mid Z = z).
\]
Since the probability mass is concentrated on a single class, we have:
\[
Y = g(Z) = g(\phi(X)) \quad \text{almost surely}.
\]
Finally, since $Y = \psi(X)$, it follows that $\psi(X) = g(\phi(X))$ almost surely, establishing (ii).

\section{Generalized merged weighted conformal prediction}

\paragraph{Weighted conformal p-value}
The weighted conformal prediction framework can also be characterized by a p-value. We define the weighted conformal p-value as follows:
\begin{equation}
\label{eq:wcp-pvalue}
    p(x,y) = \frac{ \sum_{(X_i,Y_i) \in \data } w(\phixz(X_i)) \indic{s(X_i,Y_i) \geq s(x,y)} + w(\Phi(x))    
    }{ \sum_{(X_i,Y_i) \in \data  } w(\phixz(X_i)) + w(\phixz(x)) }.
\end{equation}
As a direct consequence of Proposition~\ref{prop:wcp}, we have
\begin{equation}
\label{eq:wcp-prop}
    \Prob \lp p(\Xz,\Yz) > \alpha \rp = \Prob \lp \Yz \in \Conf{\Xz;1-\alpha}  \rp \geq 1 - \alpha. 
\end{equation}
From the above inequality, we can conclude that the weighted conformal p-variable $ p(\Xz,\Yz)$ is a valid p-variable, i.e., $\Prob \lp  p(\Xz,\Yz) \leq  \alpha \rp \leq \alpha$. For more discussions on weighted conformal p-values, see, e.g.,~\citet{barber2025unifying,jin2023model}.
For the k-th source domain, let us first define the weighted conformal p-value as follows:
\begin{equation}
\label{eq:wpk-def}
    \wpvaluek{x,y} = \frac{ \sum_{(\Xki,\Yki) \in \zkcal } w_k(\phixz(\Xki)) \indic{s(\Xki,\Yki) \geq s(x,y)} + w_k(\phixz(x))    
    }{ \sum_{(\Xki,\Yki) \in \zkcal  } w_k(\phixz(\Xki)) + w_k(\phixz(x)) }
\end{equation}
Following equation~\eqref{eq:wcp-prop}, it holds that
\begin{equation*}
   \Prob \lp  \wpvaluek{\Xz,\Yz} \leq \alpha \rp  \leq \alpha.
\end{equation*}
The weighted conformal p-values $\{\wpvaluek{\Xz,\Yz} \}$ are correlated due to their shared dependence on $(\Xz,\Yz)$. To aggregate them and form a valid prediction interval (by inverting their merged variable), the merging procedure must guarantee that the resulting variable is itself a valid p-variable. To do this, we utilize the concept of valid merging function. We define a function $g: [0,1]^{\ngroup} \times [0,1] \to [0,1]$ to be a valid merging function if, for any set of p-variables $\{p_1,\ldots, p_{\ngroup}\}$ with arbitrary dependence, it holds that
\begin{equation}
\Prob \lp g(p_1,\ldots,p_{\ngroup}, \alpha ) \leq \alpha \rp \leq \alpha, \quad \text{for all } \alpha \in [0,1].
\end{equation}
We include $\alpha$ as an argument to $g$ for generality, although the function itself may not explicitly depend on it. Subsequently, we can define merged weighted conformal p-value as follows:
\begin{equation}
\label{eq:mp-def}
    \wpvaluem{x,y; g,\alpha} =  g ( \wpvalueone{x,y}, \ldots, \wpvalueK{x,y},\alpha).
\end{equation}

Now, by inverting the merged weighted conformal p-value, we defined merged weighted conformal prediction as
\begin{equation}
    \label{eq:conf-merg}
    \Confwm{x; g, 1 - \alpha} = \lb y:  \wpvaluem{x,y; g,\alpha} > \alpha \rb.
\end{equation}
With the above notations, we are ready to state our result:
\begin{theos}
\label{theo:merge}
    Let $g$ be a valid merging function, the merged weighted conformal prediction interval defined in equation~\eqref{eq:conf-merg} satisfies
    \begin{equation}
        \Prob \lp \Yz \in \Confm{\Xz; g, 1- \alpha} \rp \geq 1 - \alpha.
    \end{equation}
\end{theos}

\noindent The procedure, specified in equation~\eqref{eq:merge-wcp}, can then be viewed as a special case of Theorem~\ref{theo:merge} by taking
\[
g(p_1, \ldots, p_{\ngroup}, \alpha) = \frac{\alpha}{K \gamma} \sum_{k\in [\ngroup]} \inde{p_k > (1 - \gamma)\alpha} 
\]
When $\gamma = (K-1)/K$, it can be verified that $g(p_1, \ldots, p_{\ngroup}, \alpha) = K \min_{k} p_k$. There are other choices of function $g$ that do not depend on $\alpha$, for instance,
see~\citet{ruger1978maximale,hommel1983tests,simes1986improved,wilson2019harmonic,vovk2020combining,vovk2022admissible} for alternatives on merging p-values under arbitrary dependence. While Theorem~\ref{theo:merge} offers a concise formulation, its practical application involves a trade-off: merging functions that ensure validity under arbitrary dependence often yield prediction intervals that are wider than necessary.

\section{Example on uncontrolled total variation distance}

Let us consider two distributions:
\[
\PXone = \mathrm{Unif}[0,1], \qquad 
\PXtwo = \mathrm{Unif}[2,3].
\]
We compare two distributions defined on $\R^{2n}$:
\begin{itemize}
    \item $\bpX$: each coordinate is drawn i.i.d. from 
    \( \PM = \tfrac{1}{2}\PXone + \tfrac{1}{2}\PXtwo \).
    \item $\btX$: exactly \( n \) coordinates in the vector are drawn from 
    \( \PXone \) and the remaining \( n \) coordinates from \( \PXtwo \).
\end{itemize}
Let us define the event
\[
A = \bigl\{ x\in\R^{2n} : \text{exactly } n \text{ of the } 2n \text{ coordinates in $x$ lie in } [0,1] \bigr\}.
\]
From the definition of $\btX$, we have $\PbtX(A) = 1$. For $\bpX$, letting $N$ denote the number of coordinates in $[0,1]$, we have \( N \sim \mathrm{Binomial}(2n, 1/2) \). Therefore,
\[
\PbpX(A) = \PbpX(N = n)
= \binom{2n}{n}\!\left(\frac{1}{2}\right)^{2n}
= \frac{\binom{2n}{n}}{4^n}.
\]
By the definition of the total variation distance, we have
\[
\mathrm{TV}(\PbpX, \PbtX)
= \sup_{B} |\PbpX(B) - \PbtX(B)|
\ge |\PbpX(A) - \PbtX(A)|
= 1 - \frac{\binom{2n}{n}}{4^n}.
\]
Using Stirling’s approximation
\[
\binom{2n}{n} \sim \frac{4^n}{\sqrt{\pi n}},
\]
we obtain, for some constant $C$ and large $n$, we have
\[
\mathrm{TV}(\PbpX, \PbtX)  \ge 1 - \frac{C}{\sqrt{\pi n}}.
\]
Thus, when $n$ is sufficiently large, we have  $\mathrm{TV}(\PbpX, \PbtX)$ is close to $1$.

\section{Proof of Theorem~\ref{theo:poolh}}

Let $N_2 = |\zcalhtwo|$. We define the estimation error $\err = \Exs | \whe(Z) - \ws (Z) |$, where the expectation is over $Z \sim \PS$ and the randomness of $\zcalhtwo$ (which determines $\{\htau_k\}$). Let $\taumin$ and $\taumax$ be the minimum and maximum mixture weights, respectively, and let $\bar{n}_k$ be the number of observations from $\PZk$ in $\zcalhtwo$. We let $\hpSe$, $\hpS$, and $p_{\star}$ denote the densities of $\sum_k \beta_k \PZk$, $\sum_k \htau_k \PZk$, and $\PS$, respectively. With a slight abuse of notation, let $p_k$ denote the density of $\PZk$ and $q$ denote the density of $\QZ$.

To begin with, we decompose the term based on a high-probability event $\calA = \left\{ \forall k: |\htau_k - \tau_k| \le \frac{\taumin}{2} \right\}$:
\begin{align*}
    \err =  \underbrace{\Exs[|\whe(Z) - \ws(Z)| \cdot \indic{\calA}]}_{\erra} + \underbrace{\Exs[|\whe(Z) - \ws(Z)| \cdot \indic{\calA^c}]}_{\errac}
\end{align*}
The event $\calA$ implies $\htau_k \ge \taumin/2$. Moreover, we assume $N_2$ is large enough such that $\epsn = 1/N_2 < \taumin/2$. We note that, on event $\calA$: $\max(\htau_k, \epsn) = \htau_k$, $\hpSe(x)=\hpS$, and $\whe = \wh$.

\paragraph{Bound on the term ($\erra$)}

Since $\whe = \wh$ on $\calA$, we have
    \[
        \erra = \Exs \left[ \left| \frac{q(Z)}{\hpS(Z)} - \frac{q(Z)}{\pS(Z)} \right| \cdot \indic{\calA} \right] = \Exs \left[ \ws(Z) \frac{|\pS(Z) - \hpS(Z)|}{\hpS(Z)} \cdot \indic{\calA} \right]
    \]
Therefore, it follows that
    \begin{align*}
        \erra & =  \Exs_{\htau} \left[ \left( \int \ws(z) \frac{\pS(z)}{\hpS(z)} |\pS(z) - \hpS(z)| dz \right) \indic{\calA} \right] \\
        & \stackrel{(i)}{\le} \Exs_{\htau} \left[ \left( \int 2\Cw  \cdot |\pS(z) - \hpS(z)| dz \right) \indic{\calA} \right] \\
        &\le 2\Cw \Exs_{\htau} \left[ \left( \int |\sum_k (\tau_k - \htau_k) p_k(z)| dz \right) \indic{\calA} \right] \\
        &\le 2\Cw \Exs_{\htau} \left[ \left( \sum_k |\tau_k - \htau_k| \int p_k(z) dz \right) \indic{\calA} \right] \\
        &\le 2\Cw \Exs_{\htau} \left[ \|\tau - \htau\|_1 \cdot \indic{\calA} \right] \leq  2\Cw \Exs_{\htau} \left[ \|\tau - \htau\|_1  \right] 
    \end{align*}
Inequality $(i)$ follows from the fact that, on event $\calA$,
\begin{align*}
    \frac{\pS(z)}{\hpS(z)} & = \frac{\sum_k \tau_k p_k(z)}{ \sum_k \htau_k p_k(z)} \leq \frac{\sum_k \tau_k p_k(z)}{ \sum_k (\tau_k - \taumin/2)) p_k(z)} \leq \frac{\sum_k \tau_k p_k(z)}{ \sum_k (\tau_k - \tau_k/2) p_k(z)} \leq 2.
\end{align*}
Now we bound $\Exs_{\htau} [ \|\tau - \htau\|_1 ]$. First we have 
    \begin{align*}
        \Exs [ \|\tau - \htau\|_1 ] &\le \Exs [ \sqrt{K} \|\tau - \htau\|_2 ] \\
        &\le \sqrt{K} \sqrt{\Exs [ \|\tau - \htau\|_2^2 ]} \quad \text{(by Jensen's inequality)}
    \end{align*}
Subsequently, we compute the expected squared $L_2$ error:
    \begin{align*}
        \Exs [ \|\tau - \htau\|_2^2 ] &= \Exs \left[ \sum_{k=1}^K (\htau_k - \tau_k)^2 \right] = \sum_{k=1}^K \Exs [ (\htau_k - \tau_k)^2 ] \\
        &= \sum_{k=1}^K \text{Var}(\htau_k) \quad \text{(since $\htau_k$ is unbiased)} \\
        &= \sum_{k=1}^K \text{Var}(\bar{n}_k / N_2) = \sum_{k=1}^K \frac{\text{Var}(\bar{n}_k)}{N_2^2} \\
        & \stackrel{(ii)}{=} \sum_{k=1}^K \frac{N_2 \tau_k (1-\tau_k)}{N_2^2} = \frac{1}{N_2} \sum_{k=1}^K (\tau_k - \tau_k^2) \\
        &= \frac{1}{N_2} \left( \sum_{k=1}^K \tau_k - \sum_{k=1}^K \tau_k^2 \right) = \frac{1 - \sum_{k=1}^K \tau_k^2}{N_2}
    \end{align*}
Equality $(ii)$ follows from $\bar{n}_k\sim \text{Binomial}(N_2, \tau_k)$. Also, we observe that the above term is maximized when $\sum \tau_k^2$ is minimized. By Cauchy-Schwarz, $1 = (\sum \tau_k)^2 \le K \sum \tau_k^2$, so $\sum \tau_k^2 \ge 1/K$. Then, we have
    \[
        \Exs [ \|\tau - \htau\|_2^2 ] \le \frac{1 - 1/K}{N_2} = \frac{K-1}{K N_2}
    \]
To conclude, we have
    \[
        \Exs [ \|\tau - \htau\|_1 ] \le \sqrt{K} \sqrt{\frac{K-1}{K N_2}} = \sqrt{K \frac{K-1}{K N_2}} = \sqrt{\frac{K-1}{N_2}}
    \]
Therefore, we have
\[
\erra \leq 2\Cw \sqrt{\frac{K-1}{N_2}}
\]

\paragraph{Bound on the second Term ($\errac$)}

Now we bound the error on the failure event $\calA^c$, using our adjusted estimator $\whe(x)$.
\[
    \errac = \Exs[|\whe(Z) - \ws(Z)| \cdot \indic{\calA^c}]
\]
By the triangle inequality, we have
\begin{align*}
    \errac &\le \Exs[(|\whe(Z)| + |\ws(Z)|) \cdot \indic{\calA^c}] \\
    &= \Exs \left[ |\whe(Z)| \cdot \indic{\calA^c} \right] + \Exs \left[ |\ws(X)| \cdot \indic{\calA^c} \right]
\end{align*}
Note that
\[
        \Exs [ |\ws(X)| \cdot \indic{\calA^c} ] \le \Cw \cdot \Prob(\calA^c)
\]
For the first term, we need a uniform bound on $\whe$. Observe that
    \begin{align*}
        \hpSe(z) &= \sum_k \beta_k p_k(z) \stackrel{(iii)}{\geq}   \sum_k \frac{\epsn}{1 + K\epsn } p_k(z) \\
        &= \frac{\epsn}{1 + K\epsn } \sum_k \tauk \frac{1}{\tauk} p_k(z) \\
        & \ge  \frac{\epsn}{(1 + K\epsn)\taumax } \sum_k \tauk  p_k(z)=  \frac{\epsn}{(1 + K\epsn)\taumax } \pS(z)
    \end{align*}
Inequality $(iii)$ follows since $\sum_k \max(\htauk, \epsn) \leq \sum_k (\htauk + \epsn)$.  Now we can bound $\whe(z)$:
    \[
        |\whe(z)| = \frac{q(z)}{\hpSe(z)} \le \frac{(1 + K\epsn)\taumax }{\epsn}\frac{q(z)}{ \pS(z)} =  \frac{(1 + K\epsn)\taumax }{\epsn}  \ws(z) \le  \frac{(1 + K\epsn)\taumax\Cw }{\epsn}
    \]
    With $\epsn = 1/N_2$, this gives $|\whe(z)| \le \Cw\taumax(K+N_2)$. By putting things together, we have
    \[
        \Exs [ |\whe(Z)| \cdot \indic{\calA^c} ] \le \Exs [ \Cw\taumax(K+N_2) \cdot \indic{\calA^c} ] = \Cw\taumax(K+N_2) \cdot\Prob(\calA^c)
    \]
Combining the two pieces for $\errac$:
\[
    \errac \le \Cw \cdot \Prob(\calA^c) + \Cw\taumax(K+N_2) \cdot \Prob(\calA^c) = \Cw\lb\taumax(K+N_2)+1\rb \cdot \Prob(\calA^c)
\]
Using the Hoeffding bound for $\Prob(\calA^c) \le 2K \exp(-N_2 \taumin^2 / 2)$:
\[
    \errac \le  \Cw\lb\taumax(K+N_2)+1\rb  \cdot 2K \exp(-N_2 \taumin^2 / 2)
\]
This term is a polynomial in $N_2$ times an exponential decay, so it is an exponentially small tail term. Combining the bounds for these two terms, we have
\[
    \err = \erra + \errac \le  2\Cw\sqrt{\frac{K-1}{N_2}}  +  \Cw\lb\taumax(K+N_2)+1\rb  \cdot 2K \exp(-N_2 \taumin^2 / 2) 
\]
Consequently, we prove that $\err = O\lp\sqrt{\frac{K}{N_2}}\rp$. Lastly, a direct application of Theorem 3 in~\cite{lei2021conformal} completes the proof.

\section{Experiment details}

\subsection{Regression task}
\subsubsection*{Sample sizes and data splitting} 
In regression task, we simulate $K \in \{5, 10\}$ distinct domains. To introduce variations in domain sizes, the total number of observations for each source group is set to $[100, 100, 100, 100, 1000] \times \lfloor d / 2 \rfloor$, where $d \in \{2, 5, 10\}$ is the covariate dimensionality. For each source domain, the generated data is split equally into a labeled training set and a calibration set (i.e., $50\%$ for training, $50\%$ for calibration). The target domain provides a set of unlabeled features; the size of this unlabeled target training set is set to half of the total combined size of all sources. Our evaluation metrics (e.g., coverage and interval width) are then computed point-wise on independently sampled target test points. One test point per replication.

\subsubsection*{Model training} 
For the predictive model, we employ a Gaussian Process regression framework. To train the model, we pool the labeled training splits from all $K$ source domains into a single, merged training dataset. We utilize a RBF kernel, and the model hyperparameters are optimized jointly on the pooled source training data.

\subsubsection*{Likelihood ratio estimation} 
We estimate the likelihood/density ratios $w$ required for weighted conformal prediction methods by casting the problem as binary classification, following~\cite{tibshirani2019conformal}. Specifically, we utilize Random Forest classifiers as follows:
\begin{itemize}
    \item \textbf{Merged WCP:} For each individual source domain $k$, a Random Forest classifier is trained to distinguish between the training data of source $k$ (assigned class $0$) and a randomly subsampled, equally-sized subset of the unlabeled target data (assigned class $1$).
    \item \textbf{Pooled WCP:} A single global Random Forest classifier is trained to distinguish the entirety of the pooled source training data (class $0$) from the complete unlabeled target data (class $1$).
\end{itemize}
The predicted class probabilities from these classifiers are subsequently converted into the likelihood ratios used to weight the non-conformity scores during the calibration phase.

\subsection{Image classification task}

\subsubsection*{Data processing and sample sizes}
To ensure consistency across the highly heterogeneous Digits-Five domains (MNIST, MNIST-M, SVHN, SYN, and USPS), we apply a standardized preprocessing pipeline to all images before model training and test. First, all images are resized to a uniform spatial resolution of $32 \times 32$ pixels. To reconcile differences between intrinsically grayscale domains (e.g., MNIST, USPS) and colored domains (e.g., SVHN, MNIST-M), we enforce a three-channel format for all inputs; naturally grayscale images are duplicated across the three RGB channels. The images are then converted into PyTorch tensors. Finally, we normalize the pixel values of each image using a mean of $0.5$ and a standard deviation of $0.5$ across all three channels, scaling the input features to a $[-1, 1]$ range.
\paragraph{Dataset size we collected}
\begin{itemize}
    \item \textbf{MNIST}: train 60,000, test 10,000
    \item \textbf{MNIST-M}: train 59,000, test 9,000
    \item \textbf{SVHN}: train 73,257, test 26,032
    \item \textbf{SYN}: train 10,000, test 2,000
    \item \textbf{USPS}: train 7,291, test 2,007
\end{itemize}
For the SYN domain, the original dataset was unavailable; therefore, we used a version obtained from Kaggle. For the uncertainty quantification experiments, we randomly sampled up to 10,000 images to construct the training split, and up to 3,000 images each for the calibration and test splits.

\subsubsection*{Model architecture}
Our model architecture is decoupled into two primary components: a multi-head feature extractor and a linear classifier. The feature extractor processes the standardized $32 \times 32$ input images using a Convolutional Neural Network (CNN) backbone comprised of three sequential convolutional blocks. Each block consists of two repeated sequences of $3 \times 3$ 2D Convolutions, 2D Batch Normalization, and ReLU activations, terminating with a $2 \times 2$ Max Pooling layer for spatial downsampling. Across the three blocks, the channel depth is progressively expanded from the 3 input channels to 64, 128, and finally 256 channels. Following the convolutional backbone, the feature maps are aggregated using a Global Adaptive Average Pooling layer and flattened into a vector with $\dz$. This vector is then routed into two distinct fully-connected heads:
\begin{itemize}
    \item \textbf{Representation Head:} A module comprising a linear layer that projects the 256-dimensional vector into a configurable latent dimension ($d_{\text{hidden}}$), followed by a ReLU activation and Dropout ($p=0.3$). 
    \item \textbf{Projection Head:} An auxiliary network used exclusively during the contrastive learning phase for domain alignment. It acts as a Multi-Layer Perceptron (MLP) mapping the latent representation $\mathbf{h}$ to a specialized projection dimension ($d_{\text{proj}}$) via two linear layers interleaved with a ReLU activation.
\end{itemize}
Here we set $d_{\text{hidden}} = \dz$ and $d_{\text{proj}} = \dz //2$. Finally, the predictor module serves as the task-specific classifier. It is implemented as a single linear layer that maps the $d_{\text{hidden}}$-dimensional latent representation directly to the 10 output logits corresponding to the digit classes.

\subsubsection*{Model training configuration}
The network training was conducted with a consistent setup across all image classification experiments. To ensure reproducibility, the random seed was fixed to 42. We utilized mini-batch gradient descent with a batch size of 64. All models were optimized using a learning rate of $5 \times 10^{-4}$ alongside a weight decay coefficient of $5 \times 10^{-4}$. The dimensionality of the latent representation was fixed at $d_{\text{hidden}} \in \{128,256,512\}$, and the projection head dimension was set to $d_{\text{proj}} = d_{\text{hidden}}//2$. Our experimental protocol evaluates two distinct training regimes:
\begin{itemize}
    \item \textbf{Base:} The unaligned baseline model was trained end-to-end for 50 epochs.
    
    \item \textbf{Base + CL:} To actively construct the domain-invariant representation required by our proposed methods, we employed a two-stage training strategy.  First, a contrastive learning phase was conducted for 50 epochs utilizing a contrastive temperature parameter of $\mathrm{temp} = 0.1$. Following this representation alignment phase, the task-specific classifier head was fine-tuned for an additional 25 epochs.
\end{itemize}
See~\cite{chen2020simple} for a demonstration of contrasive learning. For $\dz = 256$ and $\mathrm{temp} = 0.1$, we provide training trajectories for Base and Base + CL for a reference.

\begin{figure}[htbp]
    \centering
    \begin{subfigure}[b]{0.32\textwidth}
        \centering
        \includegraphics[width=\textwidth]{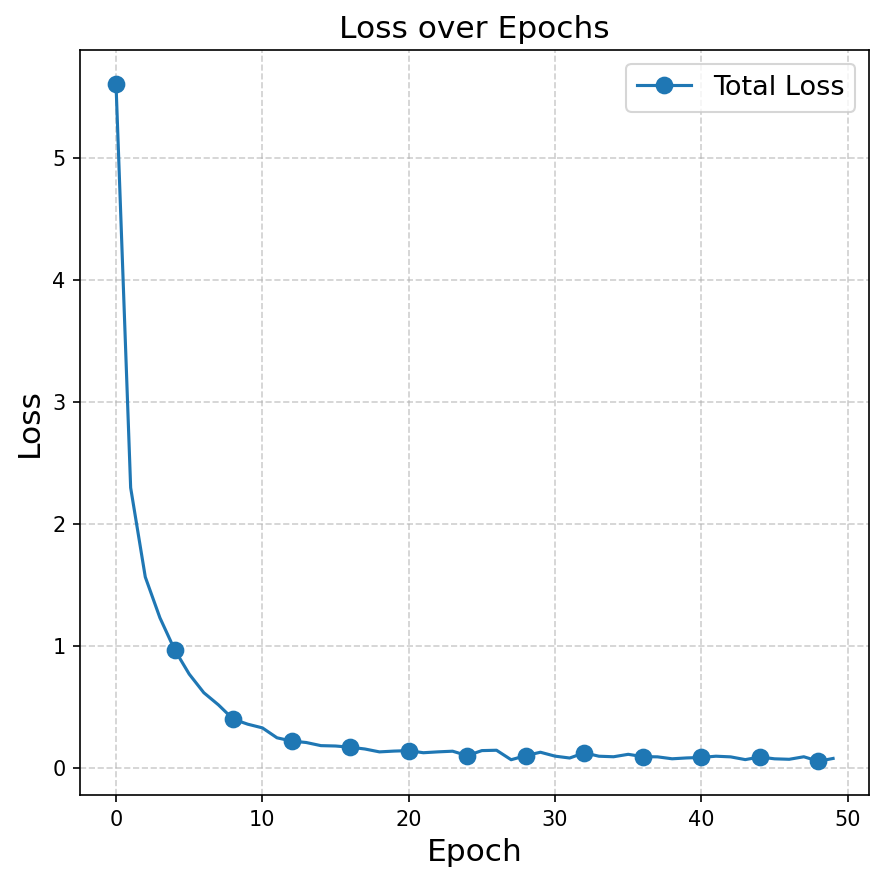}
        \caption{Training loss}
        \label{fig:base-sub1}
    \end{subfigure}
    \hfill
    \begin{subfigure}[b]{0.32\textwidth}
        \centering
        \includegraphics[width=\textwidth]{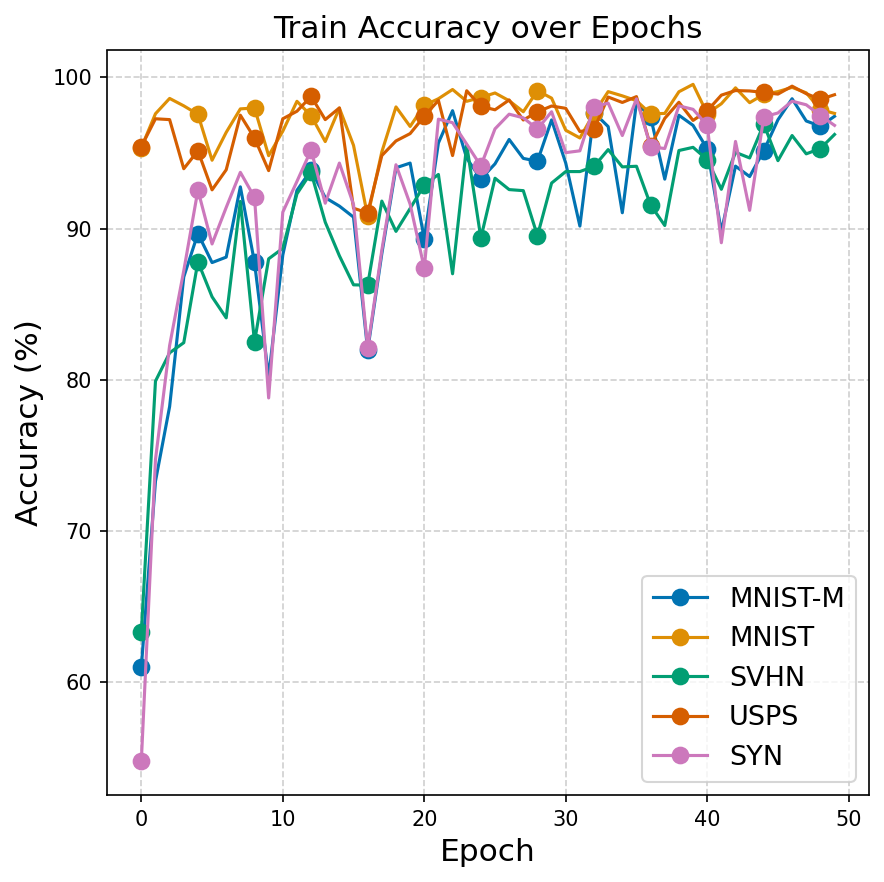}
        \caption{Training accuracy}
        \label{fig:base-sub2}
    \end{subfigure}
    \hfill 
    \begin{subfigure}[b]{0.32\textwidth}
        \centering
        \includegraphics[width=\textwidth]{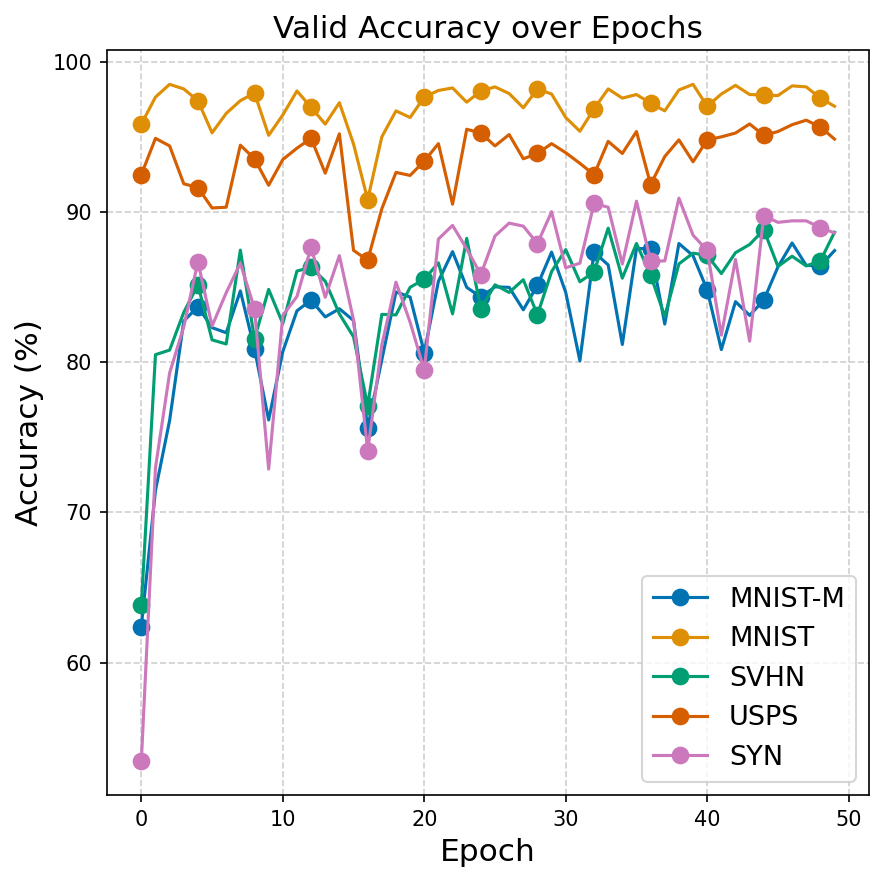}
        \caption{Validation accuracy}
        \label{fig:base-sub3}
    \end{subfigure}
    \caption{Base model training trajectory}
    \label{fig:base_three_figures}
\end{figure}

\begin{figure}[htbp]
    \centering
    \begin{subfigure}[b]{0.45\textwidth}
        \centering
        \includegraphics[width=\textwidth]{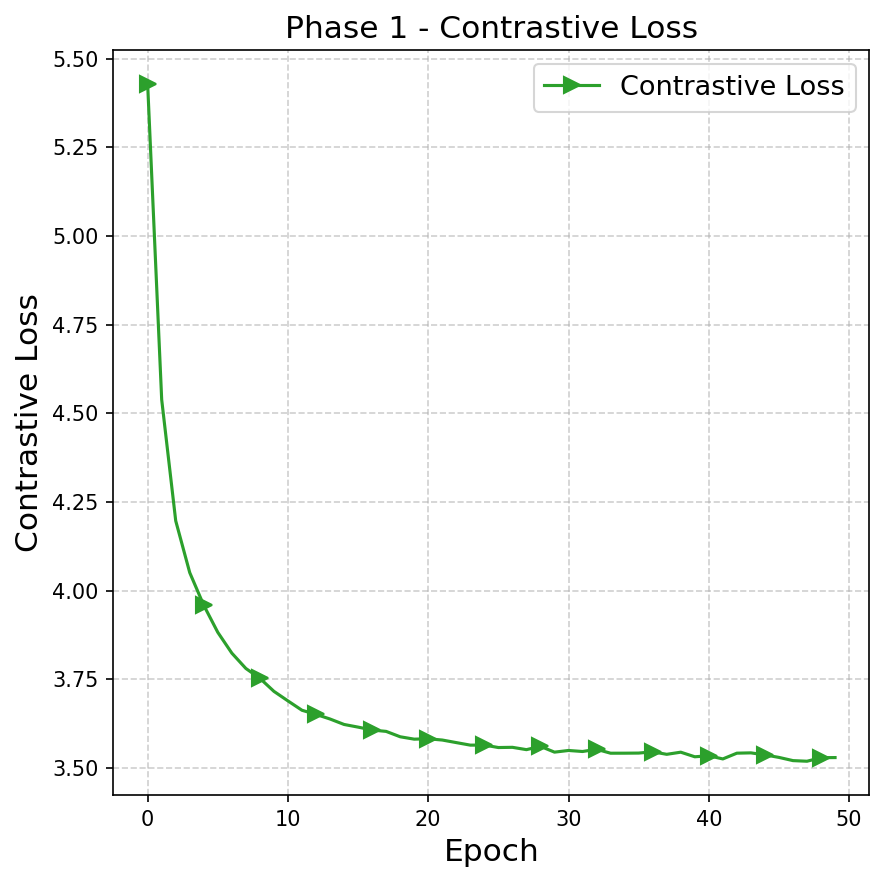}
        \caption{Training contrasive loss}
        \label{fig:cl-sub1}
    \end{subfigure}
    \hfill
    \begin{subfigure}[b]{0.45\textwidth}
        \centering
        \includegraphics[width=\textwidth]{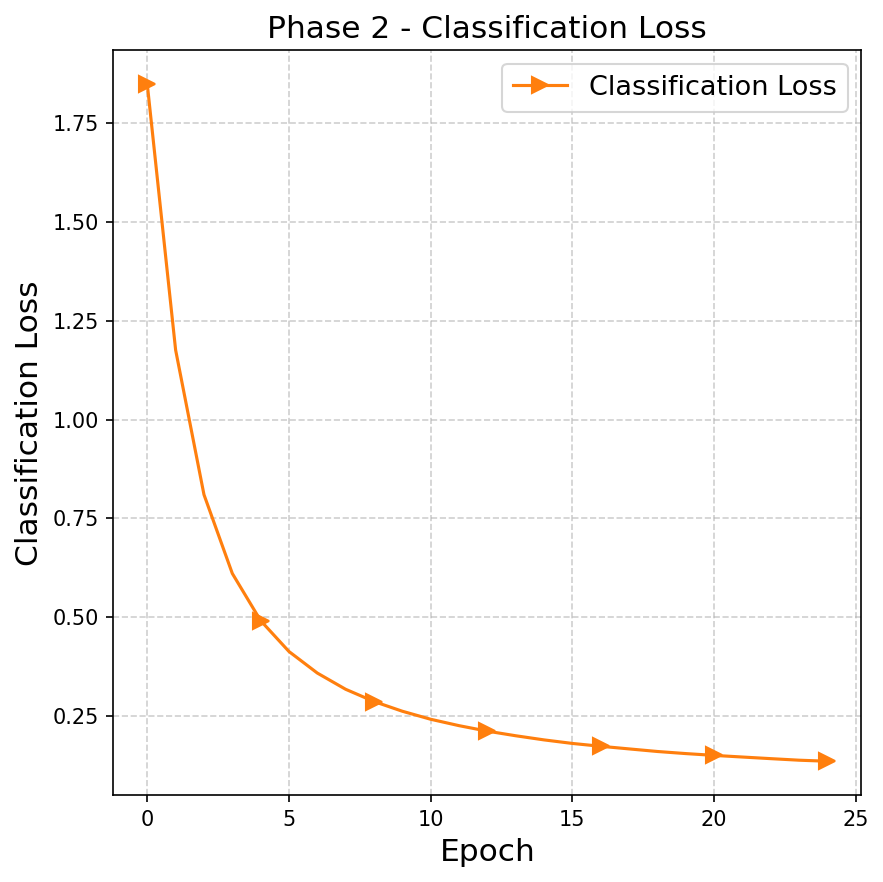}
        \caption{Training classification loss}
        \label{fig:cl-sub1}
    \end{subfigure}
    \hfill
    \begin{subfigure}[b]{0.45\textwidth}
        \centering
        \includegraphics[width=\textwidth]{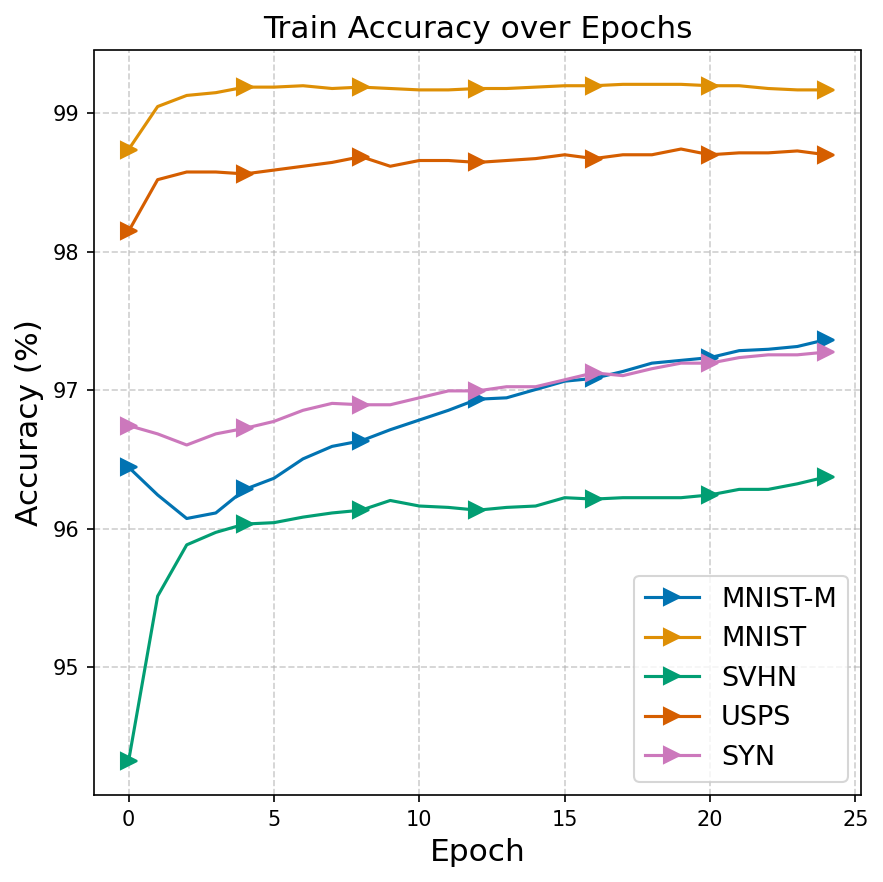}
        \caption{Training accuracy}
        \label{fig:cl-sub2}
    \end{subfigure}
    \hfill 
    \begin{subfigure}[b]{0.45\textwidth}
        \centering
        \includegraphics[width=\textwidth]{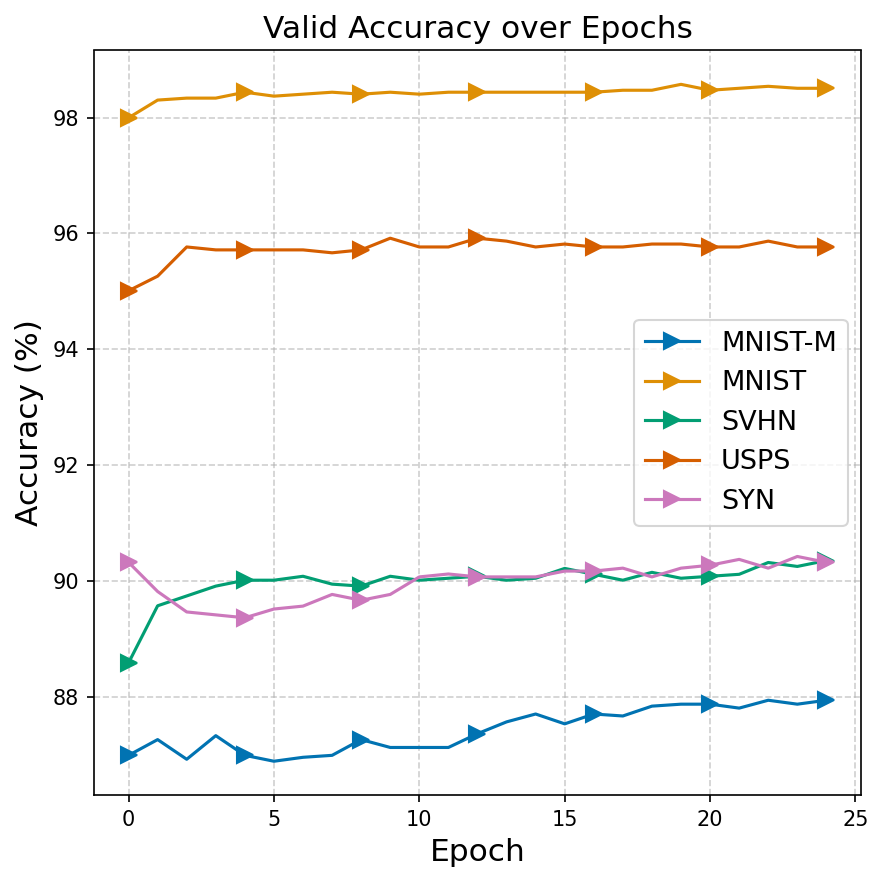}
        \caption{Validation accuracy}
        \label{fig:cl-sub3}
    \end{subfigure}
    \caption{Base + CL model training trajectory}
    \label{fig:cl_four_figures}
\end{figure}

\subsubsection*{Contrasive learning with different temperature}
In this section, we include additional experiment results with contrasive learning temperature $\{0.05,0.2\}$.
\begin{table}[!htp]
\caption{Evaluation of different latent space dimensions}
\label{tab:syn-high-dim}
\centering
\resizebox{0.95\textwidth}{!}{
\begin{tabular}{clcccccccccccc}
\toprule
& 
& \multicolumn{4}{c}{MNIST-M}
& \multicolumn{4}{c}{SVHN}
& \multicolumn{4}{c}{SYN} \\
\cmidrule(lr){3-6}
\cmidrule(lr){7-10}
\cmidrule(lr){11-14}
& 
& \multicolumn{2}{c}{$\mathrm{temp} = 0.05$}
& \multicolumn{2}{c}{$\mathrm{temp} = 0.2$}
& \multicolumn{2}{c}{$\mathrm{temp} = 0.05$}
& \multicolumn{2}{c}{$\mathrm{temp} = 0.2$}
& \multicolumn{2}{c}{$\mathrm{temp} = 0.05$}
& \multicolumn{2}{c}{$\mathrm{temp} = 0.2$} \\
\cmidrule(lr){3-4}
\cmidrule(lr){5-6}
\cmidrule(lr){7-8}
\cmidrule(lr){9-10}
\cmidrule(lr){11-12}
\cmidrule(lr){13-14}
$\mathrm{dim}(\mathcal{Z})$ & Method
& MCP & Size
& MCP &  Size
& MCP &  Size
& MCP &  Size
& MCP &  Size
& MCP &  Size \\

\midrule
\multirow{4}{*}{128}
&CP &90.1\% &1.07 &89.1\% &1.10 &92.8\% &1.11 &93.3\% &1.15 &94.0\% &1.17 &94.5\% &1.21 \\
&Pooled WCP &93.3\% &1.23 &93.4\% &1.37 &95.6\% &1.34 &95.1\% &1.36 &96.6\% &1.32 &97.3\% &1.48 \\
&Merged WCP($\gamma = 1/2$) &95.4\% &1.38 &95.6\% &1.66 &96.1\% &1.40 &96.5\% &1.69 &98.5\% &1.59 &98.7\% &1.91 \\
&Merged WCP($\gamma = (K-1)/K$) &94.8\% &1.34 &95.5\% &1.66 &96.1\% &1.40 &96.5\% &1.68 &98.7\% &1.65 &97.9\% &1.58 \\
\midrule
\multirow{4}{*}{256}
&CP &88.4\% &1.06 &87.5\% &1.00 &90.9\% &1.05 &92.4\% &1.05 &95.7\% &1.14 &95.8\% &1.09 \\
&Pooled WCP &92.3\% &1.23 &92.3\% &1.18 &95.5\% &1.29 &95.7\% &1.28 &97.4\% &1.32 &97.2\% &1.17 \\
&Merged WCP($\gamma = 1/2$) &95.8\% &1.61 &96.4\% &1.55 &97.1\% &1.58 &96.7\% &1.40 &97.9\% &1.56 &98.3\% &1.32 \\
&Merged WCP($\gamma = (K-1)/K$) &95.6\% &1.61 &97.3\% &1.71 &96.3\% &1.45 &97.9\% &1.68 &97.8\% &1.48 &98.8\% &1.50 \\
\midrule
\multirow{4}{*}{512}
& CP &88.9\% &1.02 &88.1\% &1.00 &92.8\% &1.08 &93.9\% &1.12 &94.3\% &1.12 &93.8\% &1.12 \\
&Pooled WCP &91.8\% &1.13 &93.4\% &1.25 &96.2\% &1.30 &96.5\% &1.30 &97.0\% &1.32 &96.3\% &1.29 \\
&Merged WCP($\gamma = 1/2$) &95.9\% &1.51 &97.1\% &1.75 &97.0\% &1.51 &98.1\% &1.58 &98.9\% &1.80 &98.7\% &1.72 \\
&Merged WCP($\gamma = (K-1)/K$) &95.0\% &1.37 &93.6\% &1.29 &97.3\% &1.53 &97.2\% &1.40 &98.4\% &1.57 &98.0\% &1.52 \\
\bottomrule
\end{tabular}}
\end{table}
We further evaluate additional temperatures $\mathrm{temp} \in \{0.05, 0.2\}$ and compare them with the main setting $\mathrm{temp} = 0.1$. Across all datasets and latent dimensions, the empirical coverage remains close to the nominal $95\%$ level, demonstrating that the validity of discussed approaches is robust to moderate variations in the temperature parameter. While differences in coverage and prediction set size are observed, they are generally modest and exhibit no consistent monotonic trend with respect to temperature.

\end{document}